\documentclass[a4paper]{article}
\usepackage{axodraw4j}
\usepackage{pstricks}

\usepackage{hyperref}
\usepackage{geometry}
\usepackage{amsmath}
\usepackage{authblk}
\usepackage{graphicx}
\usepackage{color}

\usepackage{ulem} %\sout{text to be striked out} 

\usepackage[T1]{fontenc}
\usepackage[sc,osf]{mathpazo}

\hypersetup{
  colorlinks = true,
  urlcolor = blue,
}

\begin{document}
%\fontsize{12}{15}
\selectfont
\title{ The Analytical One-Loop Contributions to Higgs Boson Mass
 in the Supersymmetric Standard Model with Vector-like Particles}
\author[1,2]{Tianjun Li\thanks{tli@itp.ac.cn}}
\author[3]{Wenyu Wang\thanks{wywang@mail.itp.ac.cn}}
\author[1]{Xiao-Chuan Wang\thanks{xcwang@itp.ac.cn}}
\author[3]{Zhao-Hua Xiong\thanks{xiongzh@ihep.ac.cn}}
\affil[1]{State Key Laboratory of Theoretical Physics,
Institute of Theoretical Physics, Chinese Academy of Sciences,
Beijing 100190, P. R. China}
\affil[2]{School of Physical Electronics,
University of Electronic Science and Technology of China,
Chengdu 610054, P. R. China}
\affil[3]{Institute of Theoretical Physics,
Beijing University of Technology,
Beijing 100124, P. R. China}
%\date{\today}
\date{\today}
\maketitle
%\tableofcontents

\begin{abstract}
\label{sec:abs}
In the Minimal Supersymmetric Standard Model (MSSM) with additional
vector-like particles (VLPs), we for the first time derive the particle mass spectra and 
the Feynman rules, as well as analytically calculate the one-loop contributions to the Higgs boson mass 
from the fermions and sfermions. After discussing and numerically analysing
a cases without bilinear terms and a case with a (partial) decoupling limit, we find:
(i) The corrections depend on the mass splittings between quarks and squarks
and  between vector-like fermions and their sfermions;
(ii) There exists the (partial) decoupling limit, where the VLPs decouple from the electrwoeak (EW) energy scale,
even when one of the VLPs is light around the EW scale. 
The reason is that the contributions to Higgs mass can be suppressed by 
the (or partial) decoupling effects, which
 can make the EW phenomenology very different from the MSSM;
(iii) The SM-like Higgs boson with mass around ~$125$~GeV gives strong constraints 
on the VLPs if the top squarks are around ~$1$~TeV.
Moreover, we present some numerical analyses to understand these unique features. 

\end{abstract}

\section{Introduction}
\label{sec:int}
With the observation of the Higgs boson~\cite{Aad:2012tfa, Chatrchyan:2012ufa},
the particle content of the electroweak Standard Model (SM) is confirmed 
by the experiments. In the future the main mission of
Large Hedron Collider (LHC) is to measure the interactions
involving Higgs precisely and search for the signatures of New Physics (NP).
Among all the NP models, the Minimal Supersymmetric Standard Model (MSSM)
is one of the most competitive candidates. It provides a natural
solution to the gauge hierarchy problem in the SM and realizes the gauge coupling
unification which strongly suggests Grand Unified Theories (GUTs). However,
in GUTs there generically exists the doublet-triplet splitting problem and
dimension-five proton decay problem. Fortunately, the flipped $SU(5)\times U(1)_X$ 
model~\cite{smbarr, dimitri, AEHN-0}
could elegantly solve these problems
via missing partner mechanism~\cite{AEHN-0}.
In order to explain the little hierarchy problem between
the traditional GUT scale and string scale,
one of us (TL) with Jiang and Nanopoulos proposed the
testable flipped $SU(5)\times U(1)_X$ model, dubbed as ${\cal F}$-$SU(5)$~\cite{Jiang:2009zza}, 
in which the TeV-scale vector-like particles (VLP) were introduced~\cite{Jiang:2006hf}.
Such kind of models can be constructed from the free fermionic string
constructions at the Kac-Moody level one~\cite{Antoniadis:1988tt, Lopez:1992kg}
and from the local F-theory model~\cite{Beasley:2008dc, Jiang:2009zza}.
These models are very interesting from the phenomenological point of view~\cite{Jiang:2009zza}:
the VLPs could be observed at the Large Hadron Collider (LHC),
proton decay is within the reach of the future Hyper-Kamiokande~\cite{Nakamura:2003hk} and
Deep Underground Science and Engineering Laboratory (DUSEL)~\cite{DUSEL} experiments~\cite{Li:2009fq, Li:2010dp},
the hybrid inflation could be realized naturally, the
correct cosmic primordial density fluctuations could be
got~\cite{Kyae:2005nv}, and the lightest CP-even Higgs boson
mass could be lifted~\cite{Huo:2011zt, Li:2011ab}, etc.
With no-scale boundary conditions
at $SU(5)\times{U(1)_X}$ unification scale~\cite{Cremmer:1983bf},
one of us (TL) with Maxin, Nanopoulos and Walker have
found an extraordinarily constrained ``golden point''~\cite{Li:2010ws}
and ``golden strip''~\cite{Li:2010mi} that satisfied all the latest
experimental constraints and has an imminently observable proton
decay rate~\cite{Li:2009fq}. For a review of the recent progresses,
please see Ref.~\cite{Li:2012uj}.

With the TeV-scale VLPs, the ${\cal F}$-$SU(5)$ model is
different from the MSSM at low energy.
For example, there exist non-decoupling effects in the quark and lepton sectors,
comparing with the two Higgs doublet (2HD) model.
In Ref.~\cite{Li:2012xz}, we studied the B physics processes in the
quark sector of the ${\cal F}$-$SU(5)$ model, including the quark
mass spectra, Feynman rules, new operators and Wilson coefficients.
We found that rich VLP phenomenology 
needs to be studied further, in addition to the other effects
of VLPs studied in Refs.~\cite{Huo:2011zt,Martin:2007pg,Babu:2008ge, 
Martin:2009bg,Chang:2014aaa,Nickel:2015dna}.
In this paper, we will extend our previous study
and add vector-like particle multiplets to the MSSM, dubbing as the MSSMV.
In order to show the physics of VLPs more clearly,
we concentrate on the changes compared with the MSSM.
It is well-known that only Yukawa interactions generate the masses for 
quarks and leptons in both the SM and MSSM. However, if there
are VLPs in models, the fermions could obtain their vector-like masses
via the additional bilinear mass terms. The interaction vertices will be changed
as well, so the Feynman rules and mass spectra are different from the MSSM.
We will discuss their implications on loop corrections to the Higgs mass,
(partial) decoupling suppression of fermion and sfermion sector,
and whether there exist the non-decoupling effects at the EW scale, etc.

This paper is organized as follows.
A brief description of our model, the mass matrices of all particles,
and all Feynman rules of fermions and sfermions are presented in Section~\ref{sec:model}.
Section~\ref{sec:mass} includes a complete analytical formula of
the leading order one-loop correction to the Higg mass in the MSSMV in the on-shell renormalization scheme,
and some numerical analyses. Section~\ref{sec:summary} is our summary.

\section{The MSSMV}
\label{sec:model}
\subsection{The Superpotential and Soft Terms}
\label{subsec:model}
In this subsection, we present a brief description of the MSSMV.
In addition to all the particles in the MSSM, we introduce two sets of vector-like
quarks and leptons (one set with $X$ ahead, the other set with $Y$ ahead)
and they have opposite SM quantum numbers, as given in Table~\ref{table:vlp}~\footnote{
Although we can introduce more vector-like particles 
to the Lagrangian, because they will have  similar mass spectra
and Feynman rules, we only need to add more
subscripts to the rotation matrix. Thus, we think it is enough to introduce only
one generation at current stage.}.
\begin{table}[h]
\centering
\begin{tabular}{|c|c|c|c|c|c}
\hline  SF  &  Spin 0  &  Spin \ensuremath{\frac{1}{2}} &  Generations  &
\ensuremath{(U(1)\otimes\,\text{SU}(2)\otimes\,\text{SU}(3))} \\ \hline \ensuremath{\hat{Xq}} &  \ensuremath{\tilde{Xq}} &  \ensuremath{Xq} &  1  &
\ensuremath{(\frac{1}{6},{\bf 2},{\bf 3})}   \\ \ensuremath{\hat{Xl}} &  \ensuremath{\tilde{Xl}} &  \ensuremath{Xl} &  1  &
\ensuremath{(-\frac{1}{2},{\bf 2},{\bf 1})}   \\ \ensuremath{\hat{Yq}} &  \ensuremath{\tilde{Yq}} &  \ensuremath{Yq} &  1  &
\ensuremath{(-\frac{1}{6},{\bf 2},{\bf \overline{3}})}   \\ \ensuremath{\hat{Yl}} &  \ensuremath{\tilde{Yl}} &  \ensuremath{Yl} &  1  &
\ensuremath{(\frac{1}{2},{\bf 2},{\bf 1})}   \\ \ensuremath{\hat{Xd}} & \ensuremath{\tilde{Xd_{R}^{*}}} &  \ensuremath{Xd_{R}^{*}} &  1  &
\ensuremath{(\frac{1}{3},{\bf 1},{\bf \overline{3}})}   \\ \ensuremath{\hat{Xu}} & \ensuremath{\tilde{Xu_{R}^{*}}} &  \ensuremath{Xu_{R}^{*}} &  1  &
\ensuremath{(-\frac{2}{3},{\bf 1},{\bf \overline{3}})}   \\ \ensuremath{\hat{Yd}} & \ensuremath{\tilde{Yd_{R}^{*}}} &  \ensuremath{Yd_{R}^{*}} &  1  &
\ensuremath{(-\frac{1}{3},{\bf 1},{\bf 3})}   \\ \ensuremath{\hat{Yu}} & \ensuremath{\tilde{Yu_{R}^{*}}} &  \ensuremath{Yu_{R}^{*}} &  1  &
\ensuremath{(\frac{2}{3},{\bf 1},{\bf 3})}   \\ \ensuremath{\hat{Xe}} & \ensuremath{\tilde{Xe_{R}^{*}}} &  \ensuremath{Xe_{R}^{*}} &  1  &
\ensuremath{(1,{\bf 1},{\bf 1})}   \\ \ensuremath{\hat{Ye}} &
\ensuremath{\tilde{Ye_{R}^{*}}} & \ensuremath{Ye_{R}^{*}} &  1  &
\ensuremath{(-1,{\bf 1},{\bf 1})}  \\
\hline
\end{tabular}
\caption{ The extra VLPs and their Quantum Numbers.}
\label{table:vlp}
\end{table}

It is clear that the $X$-type particles have the same quantum numbers
as the SM fermions, so we could
combine these $X$-type particles with ordinary three generation
particles to shorten the superpotential, which is given by
\begin{eqnarray}
W&=&\mu\hat{H}_{u}\hat{H}_{d}-Y_{d}\hat{d}\hat{q}\hat{H}_{d} -Y_{e}\hat{e}\hat{l}\hat{H}_{d}+Y_{u}\hat{u}\hat{q}\hat{H}_{u}\nonumber\\
&+&Y_{y_d}\hat{Yd}\hat{Yq}\hat{H}_{u}+Y_{y_e}\hat{Ye}\hat{Yl}\hat{H}_{u} -Y_{y_u}\hat{Yu}\hat{Yq}\hat{H}_{d}\nonumber\\
&+&M_{y_q}\hat{q}\hat{Yq}+M_{y_u}\hat{u}\hat{Yu}+M_{y_d}\hat{d}\hat{Yd}
+M_{y_l}\hat{l}\hat{Yl} +M_{y_e}\hat{e}\hat{Ye},
\label{super1}
\end{eqnarray}
where $\hat{H}_{d}=(H_{d}^{0}, H_{d}^{-})$ and
$\hat{H}_{u}=(H_{u}^{+}, H_{u}^{0})$ are the $SU(2)$ Higgs doublets with
hypercharges $-1/2$ and $1/2$ and have vacuum expectation values (VEVs)
$(v_{d},0)$ and $(0,v_{u})$ $(\tan\beta=v_{u}/v_{d})$, respectively.
We emphasize that in this superpotential $Y_{d},~ Y_{u}$,  and $Y_{e}$  
are $4\times4$ matrices, and $M_{y_q},~M_{y_u},~M_{y_d},~M_{y_l}$ and
 $M_{y_e}$ are $4\times1$ matrices.
So in Eq.~(\ref{super1}) the first line are the superpotential which is the
same as the MSSM in format. The second line are terms involving  $Y$-type VLPs
and the third line are the bilinear terms with mass dimensional matrix
$M_{y_q},~M_{y_u},~M_{y_d},~M_{y_l}$, and $M_{y_e}$ as input parameters. Compared with
the MSSM, they are new terms in the MSSMV.
We will concentrate on the implications of these terms
in the following Section. The package SARAH$4$~\cite{Staub:2013tta} is used.

The next ingredient is the supersymmetry breaking soft terms.
The gaugino masses are
\begin{eqnarray}
\label{LSBlambda}
-\mathcal{L}_{SB,\lambda}=\frac{1}{2}\Bigl(M_{1}\lambda_{\tilde{B}}^{2}+M_{2}\lambda_{\tilde{W}}^{2}
+M_{3}\lambda_{\tilde{g}}\lambda_{\tilde{g}}+\mbox{h.c.}\Bigr),
\end{eqnarray} 
 the scalar masses are 
\begin{eqnarray}
\label{LSBphi}
-\mathcal{L}_{SB,\phi}&=&m_{H_{d}}^{2}|H_{d}^{0}|^{2}
+ m_{H_{d}}^{2}|H_{d}^{-}|^{2} + m_{H_{u}}^{2}|H_{u}^{0}|^{2}
+ m_{H_{u}}^{2}|H_{u}^{+}|^{2}\nonumber\\
&+& m_{q}^{2}|\tilde{u}_{L}|^2 + m_{u}^{2}|\tilde{u}_{R}|^2
+ m_{q}^{2}|\tilde{d}_{L}|^2 + m_{d}^{2}|\tilde{d}_{R}|^2
+ m_{l}^{2}|\tilde{e}_{L}|^2 + m_{e}^{2}|\tilde{e}_{R}|^2\nonumber\\
&+& m_{Yq}^{2}|\tilde{Yu}_{L}|^2 + m_{Yu}^{2}|\tilde{Yu}_{R}|^2
+ m_{Yq}^{2}|\tilde{Yd}_{L}|^2 + m_{Yd}^{2}|\tilde{Yd}_{R}|^2\nonumber\\
&+& m_{Yl}^{2}|\tilde{Ye}_{L}|^2 + m_{Ye}^{2}|\tilde{Ye}_{R}|^2~,~
\end{eqnarray}
and the trilinear soft terms are 
\begin{eqnarray}
\label{LSBW}
-\mathcal{L}_{SB,W}&=&-H_{d}^{0}H_{u}^{0}B_{\mu}\mu+H_{d}^{-}H_{u}^{+}B_{\mu}\mu\nonumber\\
&+&H_{d}^{0}\tilde{d}_{R}^{*}\tilde{d}_{L}A_{d}
-H_{d}^{-}\tilde{d}_{R}^{*}\tilde{u}_{L}A_{d}
+H_{d}^{0}\tilde{e}_{R}^{*}\tilde{e}_{L}A_{e}
-H_{d}^{-}\tilde{e}_{R}^{*}\tilde{\nu}_{L}A_{e}
+H_{u}^{0}\tilde{u}_{R}^{*}\tilde{u}_{L}A_{u}
-H_{u}^{+}\tilde{u}_{R}^{*}\tilde{d}_{L}A_{u} \nonumber\\
&+&H_{u}^{0}\tilde{Yd}_{R}^{*}\tilde{Yd}_{L}A_{y_d}
-H_{u}^{+}\tilde{Yd}_{R}^{*}\tilde{Yu}_{L}A_{y_d}
+H_{u}^{0}\tilde{Ye}_{L}\tilde{Ye}_{R}^{*}A_{y_e}
-H_{u}^{+}\tilde{Y\nu}_{L}\tilde{Ye}_{R}^{*}A_{y_e}\nonumber\\
&+&H_{d}^{0}\tilde{Yu}_{R}^{*}\tilde{Yu}_{L}A_{y_u}
-H_{d}^{-}\tilde{Yu}_{R}^{*}\tilde{Yd}_{L}A_{y_u}
+\tilde{Xu}_{L}\tilde{Yu}_{L}B_{My_q} M_{y_q}
+\tilde{Xu}_{R}^{*}\tilde{Yu}_{R}^{*}B_{My_u}M_{y_u}\nonumber\\
&-&\tilde{Xd}_{L}\tilde{Yd}_{L}B_{My_q} M_{y_q}
+\tilde{Xd}_{R}^{*}\tilde{Yd}_{R}^{*}B_{My_d}M_{y_d}
-\tilde{Ye}_{L}\tilde{Xe}_{L}B_{My_l}
+\tilde{Ye}_{R}^{*}\tilde{Xe}_{R}^{*}B_{My_e} M_{y_e} +\mbox{h.c.}.
\end{eqnarray}
The Lagrangian characterizing the fermion, sfermion and gaugino interactions is
\begin{eqnarray}
\label{LFSFG}
-\mathcal{L}_{F,SF,G}=-\left[\frac{g_{2}}{\sqrt{2}}\left(\bar{u}_{L}^{i},~\bar{d}_{L}^{i}\right)
\lambda_{\tilde{W},a}\sigma_{ij}^{a}\left(\begin{array}{c}
\tilde{u}_{L}^{j}\\
\tilde{d}_{L}^{j}
\end{array}\right)+\frac{2}{3}\frac{g_{1}}{\sqrt{2}}\bar{u}_{R}^{i}\lambda_{\tilde{B}}\delta_{ij}\tilde{u}_{R}^{j}
-\frac{1}{3}g_{1}\sqrt{2}\bar{d}_{R}^{i}\lambda_{\tilde{B}}\delta_{ij}\tilde{d}_{R}^{j}
+\mbox{h.c.}\right]&&\nonumber\\
-\left[\frac{g_{2}}{\sqrt{2}}\left(
\overline{Yd}_{L},~\overline{Yu}_{L}\right)\lambda_{\tilde{W},a}\sigma^{a}\left(\begin{array}{c}
\tilde{Yd}_{L}\\
\tilde{Yu}_{L}
\end{array}\right)+\frac{1}{3}\frac{g_{1}}{\sqrt{2}}\overline{Yd}_{R}\lambda_{\tilde{B}}\tilde{Yd}_{R}^{j}-\frac{2}{3}g_{1}
\sqrt{2}\overline{Yu}_{R}\lambda_{\tilde{B}}\tilde{Yu}_{R}^{j}+\mbox{h.c.}\right],&&\;
\end{eqnarray}
in which the terms in the first line are same as the MSSM in form,
and the second line has the terms introduced by the $Y$-type VLPs 
which are the new terms in the MSSMV.

Generally the constants $\mu$, $B_\mu$, Yukawa matrices, squark and
gaugino masses, bilinear matrices and the trilinear soft terms
may be complex. One can eliminate the non-physical degrees of freedom
by redefining the global phases of the fields. For example,
the Higgs multiplets could be redefined so that the
constant $B_\mu$ becomes a real number.
Then the minimization equations for the VEVs of the Higgs fields will only involve real parameters.
After the proper redefinition of the parameters, 
the rotation matrices of quarks and leptons 
still remain in the Lagrangian, leaving only a Kobayashi-Maskawa like matrix and
tree-level Flavor-Changing Neutral Currents (FCNC) interactions.

\subsection{The Particle Spectra}

To obtain the physical spectra of particles present in the MSSMV
one should carry out the standard procedure of gauge symmetry breaking
via the VEVs of the neutral Higgs fields and calculate the
eigenstates of the mass matrices for all fields.
The Higgs and gaugino sectors in the MSSMV at tree level is the same as the MSSM, 
whereas only the fermion and sfermion sectors are different.
The mass spectra of fermions and sfermions are given as follows
\begin{itemize}
\item Fermion Mass Matrices
\end{itemize}
With the participation of VLPs,
the $3\times 3$ quark/lepton mass matrices become the $5\times 5$ mass matrices
 \begin{eqnarray}
m_{d}=\left(\begin{array}{cc}
\frac{1}{\sqrt{2}}v_{d}Y_{d}^{T} & -M_{y_q,{o_{1}}}^{T}\\
M_{y_d,{p_{1}}} & \frac{1}{\sqrt{2}}v_{u}Y_{y_d}
\end{array}\right),~~
m_{u}=\left(\begin{array}{cc}
\frac{1}{\sqrt{2}}v_{u}Y_{u}^{T} & M_{y_q,{o_{1}}}^{T}\\
M_{y_u,{p_{1}}} & \frac{1}{\sqrt{2}}v_{d}Y_{y_u}
\end{array}\right),~~
m_{e}=\left(\begin{array}{cc}
\frac{1}{\sqrt{2}}v_{d}Y_{e}^{T} & -M_{y_l,{o_{1}}}^{T}\\
M_{y_e,{p_{1}}} & \frac{1}{\sqrt{2}}v_{u}Y_{y_e}
\end{array}\right).
\label{fermmass}
\end{eqnarray}

The $X$-type fermions can be considered as the fourth generation,
and the mixings between the $X$-type and ordinary fermions give
 the upper left $4\times 4$ elements of the matrices.
The $Y$-type fermions give a different form in comparison with the fifth generation. 
Especially, the bilinear mass parameters $M_{yq}$ connect the mass matrices of up-type and down-type quarks, 
so their diagonaliztions are not as simple as the MSSM.
We use the following convention for the diagonalization
 \begin{eqnarray}
U_{L}^{f\dagger}m_{f}U_{R}^{f}=m_{f}^{diag.},
 \end{eqnarray}
in which  $U_{L,R}^{f}$ represent the rotation matrices and
the superscript $f$ represents $d,~u$, and $e$.
\begin{itemize}
\item Sfermion Mass Matrices
\end{itemize}
The squark and slepton mass matrices are extended from $6\times6$ to
$10\times10$. The mass matrices for up-type squarks are given by
\begin{eqnarray}
m_{{\tilde{u}}}^{2}=\left(\begin{array}{cccc}
m_{\tilde{u}_{L}\tilde{u}_{L}^{*}} & m_{\tilde{u}_{R}\tilde{u}_{L}^{*}} & B_{{My_q},{o_{1}}}^{*}M_{y_q,{o_{1}}}
 & m_{\tilde{Yu}_{R}^{*}\tilde{u}_{L}^{*}}\\
m_{\tilde{u}_{L}\tilde{u}_{R}^{*}} & m_{\tilde{u}_{R}\tilde{u}_{R}^{*}} & m_{\tilde{Yu}_{L}^{*}\tilde{u}_{R}^{*}}
& B_{My_u,{o_{2}}}M_{y_u,{o_{2}}}\\
B_{My_q,{p_{1}}}M_{y_q,{p_{1}}} & m_{\tilde{u}_{R}\tilde{Yu}_{L}}
& m_{\tilde{Yu}_{L}^{*}\tilde{Yu}_{L}} & m_{\tilde{Yu}_{R}^{*}\tilde{Yu}_{L}}\\
m_{\tilde{u}_{L}\tilde{Yu}_{R}} & B_{{My_u},{p_{2}}}^{*}M_{y_u,{p_{2}}}
& m_{\tilde{Yu}_{L}^{*}\tilde{Yu}_{R}} & m_{\tilde{Yu}_{R}^{*}\tilde{Yu}_{R}}
\end{array}\right)\label{sumasss}
\end{eqnarray}
with
\begin{eqnarray}
&&m_{\tilde{u}_{L}\tilde{u}_{L}^{*}}=-\frac{1}{24}\Big(-3g_{2}^{2}+g_{1}^{2}\Big)\Big(-v_{u}^{2}+v_{d}^{2}\Big)+\frac{1}{2}
\Big(2\Big(M_{y_q,{o_{1}}}M_{y_q,{p_{1}}}+m_{q}^{2}\Big)+v_{u}^{2}{Y_{u}^{\dagger}Y_{u}}\Big),\nonumber\\
&&m_{\tilde{u}_{R}\tilde{u}_{L}^{*}}=\frac{1}{\sqrt{2}}\Big(-v_{d}Y_{u}^{\dagger}\mu+v_{u}A_{u}^{\dagger}\Big),\nonumber\\
&&m_{\tilde{Yu}_{R}^{*}\tilde{u}_{L}^{*}}=\frac{1}{\sqrt{2}}\Big(v_{d}Y_{y_u}M_{y_q,{o_{1}}}+v_{u}Y_{u,{o_{1}}}^{*}M_{y_u}\Big),\nonumber\\
&&m_{\tilde{u}_{L}\tilde{u}_{R}^{*}}=\frac{1}{\sqrt{2}}\Big(-v_{d}Y_{u}\mu^{*}+v_{u}A_{u}\Big),\nonumber\\
&&m_{\tilde{u}_{R}\tilde{u}_{R}^{*}}=\frac{1}{2}\Big(2\Big(M_{y_u,{o_{2}}}M_{y_u,{p_{2}}}+m_{u}^{2}\Big)
+v_{u}^{2}{Y_{u}Y_{u}^{\dagger}}\Big)+\frac{1}{6}g_{1}^{2}\Big(-v_{u}^{2}+v_{d}^{2}\Big),\nonumber\\
&&m_{\tilde{Yu}_{L}^{*}\tilde{u}_{R}^{*}}=\frac{1}{\sqrt{2}}\Big(v_{d}Y_{y_u}M_{y_u,{o_{2}}}+v_{u}M_{y_q}Y_{u,{o_{2}}}\Big),\nonumber\\
&&m_{\tilde{u}_{R}\tilde{Yu}_{L}}=\frac{1}{\sqrt{2}}\Big(v_{d}Y_{y_u}M_{y_u,{p_{2}}}+v_{u}M_{y_q}Y_{u,{p_{2}}}^{*}\Big),\\
&&m_{\tilde{Yu}_{L}^{*}\tilde{Yu}_{L}}=\frac{1}{24}\Big(-3g_{2}^{2}+g_{1}^{2}\Big)\Big(-v_{u}^{2}+v_{d}^{2}\Big)+
\frac{1}{2}\Big(2\Big(m_{Yq}^{2}+M_{y_q}^{2}\Big)+v_{d}^{2}Y_{y_u}^{2}\Big),\nonumber\\
&&m_{\tilde{Yu}_{R}^{*}\tilde{Yu}_{L}}=\frac{1}{\sqrt{2}}\Big(v_{d}A_{y_u}-v_{u}Y_{y_u}\mu^{*}\Big),\nonumber\\
&&m_{\tilde{u}_{L}\tilde{Yu}_{R}}=\frac{1}{\sqrt{2}}\Big(v_{d}Y_{y_u}M_{y_q,{p_{1}}}+v_{u}Y_{u,{p_{1}}}M_{y_u}\Big),\nonumber\\
&&m_{\tilde{Yu}_{L}^{*}\tilde{Yu}_{R}}=\frac{1}{\sqrt{2}}\Big(v_{d}A_{y_u}^{*}-v_{u}Y_{y_u}\mu\Big),\nonumber\\
&&m_{\tilde{Yu}_{R}^{*}\tilde{Yu}_{R}}=\frac{1}{2}\Big(2\Big(m_{Yu}^{2}+M_{y_u}^{2}\Big)
+v_{d}^{2}Y_{y_u}^{2}\Big)+\frac{1}{6}g_{1}^{2}\Big(-v_{d}^{2}+v_{u}^{2}\Big)\nonumber.
\end{eqnarray}
In addition to the elements like the MSSM, $m_{\tilde{Yu}_{R}^{*}\tilde{u}_{L}^{*}},~M_{y_q,{o_{1}}}$ 
and the similar terms arising from the F-term potential of the bilinear terms are the special terms,
and the $B_{{My_q},{o_{1}}}$ and similar terms coming from the soft terms, 
as mentioned in subsection~\ref{subsec:model}, are also the special terms.
The mass matrices for down-type squarks and the sleptons are similar. We use the following convention
\begin{eqnarray}
Z^{F,\dagger}m_{\tilde{f}}^{2}Z^{F}=m_{2,\tilde{f}}^{diag.}
\end{eqnarray} for the diagonalization,
where $f=~d,~u,~e$, $F=~D,~U,~E$, and $Z^{F}$ are the rotation matrices.

\subsection{The Feynman Rules}

With the Lagrangians and mass spectra given above,
we can derive all Feynman rules of the sfermion sectors.
Due to the extra VLPs, the interactions
involving quarks/squarks and leptons/sleptons will be different
from the MSSM.  We list all the Feynman rules
that are different from the MSSM in the Appendix,
and just give a few comments here:
\begin{enumerate}
\item {\em The tree-level FCNC processes in the fermion sectors.} It is well-known that
both in the SM and in the MSSM, there is no tree-level FCNC process for
quarks and leptons. However, with the VLPs, such processes
will emerge. For example, the Feynman rule for the $\bar{d}_i d_j Z_\mu$ vertex is
\begin{eqnarray}
\label{ddz}
&&\frac{ie}{\sin2\Theta_{W}}\gamma_{\mu}\Bigl[-\frac{2}{3}\sin^{2}\Theta_{W}\delta_{ij}
+\left(\delta_{ij}-U_{L,{j5}}^{d,*}U_{L,{i5}}^{d}\right)P_{L}+U_{R,{i5}}^{d,*}U_{R,{j5}}^{d}P_{R}\Bigl]~,
\end{eqnarray}
and that for the $\bar d_i d_j h_k$ vertex is
\begin{eqnarray}
\label{hdd}
&& -i\frac{1}{\sqrt{2}}\Big[\big(U_{R,{ia}}^{d,*}Y_{d,{ab}}U_{L,{jb}}^{d,*}
Z_{{k1}}^{H}+Y_{y_d}U_{L,{j5}}^{d,*}U_{R,{i5}}^{d,*}Z_{{k2}}^{H}\big)P_L\,\,  \nonumber\\
&& +\big(U_{R,{ja}}^{d}Y_{d,{ab}}^{*}U_{L,{ib}}^{d}Z_{{k1}}^{H}
+Y_{y_d}U_{L,{i5}}^{d}U_{R,{j5}}^{d}Z_{{k2}}^{H}\big)P_R\Big]~,
\end{eqnarray}
where the interaction state subscripts $a$ and $b$ run from 1 to 4.
If we set the subscripts $i\ne j$, we can get the tree-level FCNC interations. Generally
speaking, such interactions should be much smaller than the other SM interactions.
\item {\em The deviation from unitarity.}
Accompanying with the tree-lever FCNC interactions, the vertex which involves the rotation
matrix will be non-unitarity. For example, the $\bar{u}_i d_j W^-_\mu$ vertex is
\begin{eqnarray}
-i\frac{g_{2}}{\sqrt{2}}\gamma_{\mu}\Bigl[U_{L,{ja}}^{d,*}U_{L,{ia}}^{u}P_{L}
-U_{R,{i5}}^{u,*}U_{R,{j5}}^{d}P_{R}\Bigl].
\end{eqnarray}
Since the interaction state subscript $a$ runs from 1 to 4,
it makes the summation non-unitarity.
In the SM, the quark mixings are described by a unitarity CKM matrix
$$V_{\rm CKM}=\sum_{a=1}^{3}U_{L,{ja}}^{d,*}U_{L,{ia}}^{u}\,.$$
Such deviation commonly emerges in every interaction listed in the Appendix.
Of course, the deviation form the unitarity should be very small. Otherwise,
it will be excluded by the current experimental limits.
These tree-level FCNC terms of Eqs. (\ref{ddz}) and (\ref{hdd}), which are called as "tail terms",
lead to rich phenomenology for low energy processes \cite{Li:2012xz}.
\item {\em The interactions from F-term potential of the bilinear terms.}  The tree-level FCNC processes
and deviations from the unitarity commonly exist in the SUSY particle sector (shown
in the Appendix). In addition, from Eq.~(\ref{sumasss}) one can easily find that the
mass contents and interactions form F-term potential are also different from the MSSM.
For example, in the $\tilde{u}_i \tilde u_j h_i$ vertex, there exist the following terms 
\begin{eqnarray}
\cdots+6\sqrt{2}
\Bigl(Y_{u,{ba}}^{*}M_{y_q,{a}}Z_{j4+b}^{U,*}Z_{{k9}}^{U}
+Y_{u,{ab}}^{*}M_{y_u,{a}}Z_{j10}^{U,*}Z_{{kb}}^{U}\Bigr)Z_{{i2}}^{H}+\cdots,
\end{eqnarray}
which come form the F-term potential, namely the bilinear terms in the superpotential.
\end{enumerate}

\section{The Leading Radiative Corrections to Higgs Boson Mass in the MSSMV}
\label{sec:mass}

As mentioned above, the tree-level FCNC processes and deviationd from the unitarity in the MSSMV
should be much smaller, so their radiative corrections can be neglected in most
processes. The new terms from the F-term potential of bilinear terms
are the first things which should be checked. Their one-loop contributions to the 
Higgs boson mass can show the dominant difference between
the MSSM and the MSSMV. In this Section, for simplicity, we choose only
one pair of the VLPs, namely, only one $X$ and one $Y$ for our study.
Note that the contribution from $X_u$ and $X_q$ particles will be exactly
the same as top quark in format sicne they have the same quantum numbers.
In the following we will study the difference  between the
contributions from $X$ and $Y$ particles.
The superpotential containing the bilinear terms is
\begin{eqnarray}
W&=&\mu\hat{H}_{u}\hat{H}_{d}+Y_{x_u}\hat X_u\hat X_q \hat{H}_{u}
  -Y_{y_u}\hat Y_u\hat Y_q \hat{H}_{d}\nonumber\\
 && +M_{y_q}\hat X_q\hat Y_q + M_{y_u} \hat X_u \hat Y_u \label{sup-sip}
\end{eqnarray}
In this simplified model, for the quark/squark sector there are two quarks
(labeled as $m_i,~~i =1,~2$ )
and four  squarks (labeled as $m_{\tilde i},~~i =~1,~2,~3,~4$ ).
 The mass matrices of Eqs.~(\ref{fermmass}) and (\ref{sumasss})
are simplified as well.

The radiative corrections to Higgs boson mass have been studied in Ref.~\cite{Huo:2011zt}
which used the effective potential method. A more detailed study should be
on-shell renomalization. In this work we follow Ref.~\cite{Dabelstein:1994hb},
using on-shell renomalization to obtain  the analytical results of
the leading-order radiative corrections to the Higgs boson mass, which are
\begin{eqnarray}
\label{equmh1}
m_{h,H}^{2} &=& \frac{m_{A}^{2}+m_{Z}^{2}+w+\sigma}{2}
\mp  \biggl[\frac{(m_{A}^{2}+m_{Z}^{2})^{2}+(w-\sigma)^{2}}{4}
-m_{A}^{2}m_{Z}^{2}\cos^{2}2\beta\nonumber\\
&+& \frac{(w-\sigma)\cos2\beta}{2}(m_{A}^{2}-m_{Z}^{2})
-\frac{\lambda\sin2\beta}{2}(m_{A}^{2}+m_{Z}^{2})
+\frac{\lambda^{2}}{4}\biggr]^{\frac{1}{2}}~,
\end{eqnarray}
where $h$ is the SM-like Higgs, and $H$ is the 
CP-even heavy Higgs in the MSSMV. In this work, we concentrate on the radiative
corrections to the SM-like Higgs boson mass.
$\omega,~\lambda,~\sigma$ are obtained from the self-energy diagrams for
the Higgs and gauge bosons.
Neglecting  the tree-level FCNC processes and deviations from unitarity, we get
the leading terms of $\omega,~\lambda,~\sigma$  in the MSSMV 
\begin{eqnarray}
\omega &=& \frac{N_{c}G_{F}v^{2}}{4\sqrt{2}\pi^{2}}\biggl\{2Y_{x_u}^{2}
\sum_{i,j}U_{L,1i}^{u\dagger}U_{R,1j}^{u}U_{L,1i}^{u\dagger}U_{R,1j}^{u}m_{i}m_{j}
B_{0}(k^{2},m_{i}^{2},m_{j}^{2})\nonumber\\
&+&Y_{x_u}^{2}\left[\frac{1}{2}Y^2_{x_u}v^2_{u}(X_{1111}+X_{2222})
+M_{y_q}^{2}X_{2323}+M_{y_u}^{2}X_{1414}\right]\nonumber\\
&+&2Y_{x_u}^{2}A_{x_u}\left[\frac{1}{\sqrt{2}}Y_{x_u}v_{u}(X_{2122}+X_{2111})
+M_{y_q}X_{2123}+M_{y_u}X_{2114}\right]\nonumber\\
&+&Y_{x_u}^{2}\left[A_{x_u}^{2}X_{2121}
+{\sqrt{2}}Y_{x_u}v_{u}(M_{y_q}X_{2322}+M_{y_u}X_{1411})\right]\nonumber\\
&+&\left[Y_{y_u}^{2}\mu^{2}X_{4343}-2Y_{x_u}Y_{y_u}\mu(M_{y_q}X_{4323}
-M_{y_u}X_{4314})\right]\biggr\},\label{omg-all}
\end{eqnarray}
\begin{eqnarray}
\sigma &=& \frac{N_{c}G_{F}v^{2}}{4\sqrt{2}\pi^{2}}\biggl\{2Y_{y_u}^{2}
\sum_{i,j}U_{L,2i}^{u\dagger}U_{R,2j}^{u}U_{L,2i}^{u\dagger}U_{R,2j}^{u}m_{i}m_{j}
B_{0}(k^{2},m_{i}^{2},m_{j}^{2})\nonumber\\
&+&Y_{y_u}^{2}\left[\frac{1}{2}Y^2_{y_u}v^2_{d}(X_{3333}+X_{4444})
+M_{y_u}^{2}X_{2323}+M_{y_q}^{2}X_{1414}\right]\nonumber\\
&+&2Y_{y_u}^{2}A_{y_u}\left[\frac{1}{\sqrt{2}}Y_{y_u}v_{d}(X_{4344}+X_{4333})
+M_{y_u}X_{4323}+M_{y_q}X_{4314}\right]\nonumber\\
&+&Y_{y_u}^{2}\left[A_{y_u}^{2}X_{4343}
+{\sqrt{2}}Y_{y_u}v_{d}(M_{y_u}X_{2333}+M_{y_q}X_{1444})\right]\nonumber\\
&+&\left[Y_{x_u}^{2}\mu^{2}X_{2121}-2Y_{x_u}Y_{y_u}\mu(M_{y_u}X_{2123}
+M_{yq}X_{2114})\right]\biggr\},\label{sig-all}
\end{eqnarray}
\begin{eqnarray}
\lambda &=& \frac{N_{C}G_{F}v^{2}}{2\sqrt{2}\pi^{2}}\biggl\{
\mu\left[\frac{1}{\sqrt{2}}Y^3_{x_u}v_{u}(X_{2122}+X_{2111})
+\frac{1}{\sqrt{2}}Y^3_{y_u}v_{d}(X_{4344}+X_{4333})\right]\nonumber\\
&+&\mu\left[Y_{x_u}^{2}(M_{y_q}X_{2123}+M_{y_u}X_{2114})
+Y_{y_u}^{2}(M_{y_u}X_{4323}+M_{y_q}X_{4314})\right]\nonumber\\
&+&\mu(Y_{x_u}^{2}A_{x_u}X_{2121}+Y_{y_u}^{2}A_{y_u}X_{4343})\nonumber\\
&-&Y_{x_u}Y_{y_u}\left[\frac{1}{\sqrt{2}}Y_{x_u}v_{u}(M_{y_u}X_{2322}+M_{y_q}X_{1411})
+\frac{1}{\sqrt{2}}Y_{y_u}v_{d}(M_{y_q}X_{2333}+M_{y_u}X_{1444})\right]\nonumber\\
&-&Y_{x_u}Y_{y_u}\left[A_{x_u}(M_{y_u}X_{2123}+M_{y_q}X_{2114})+A_{y_u}(M_{y_q}X_{4323}+M_{y_u}X_{4314})\right]\nonumber\\
&-&Y_{x_u}Y_{y_u}M_{y_q}M_{y_u}(X_{2233}+X_{1144})\biggr\},\label{lam-all}
\end{eqnarray}
where $X_{klmn}$ is defined as
\begin{eqnarray}
X_{klmn}\equiv \sum_{\tilde i,\tilde j}Z_{\tilde k\tilde i}^{U}Z_{\tilde i\tilde l}^{UT}
Z_{\tilde m\tilde j}^{U}Z_{\tilde j\tilde n}^{UT}
\times B_{0}(\mu^{2},m_{\tilde{i}}^{2},m_{\tilde{j}}^{2})~,~
\end{eqnarray}
which come from the summation of the propagators in the loop. The convergent loop function is
\begin{eqnarray}
B_{0}(\mu^{2},x^{2},y^{2})=\begin{cases}
\begin{array}{cc}\log\frac{\mu^{2}}{y^{2}}+1+\frac{x^{2}/y^{2}}{1-x^{2}/y^{2}}\log\frac{x^{2}}{y^{2}} & \mbox{if}\;
 x\ne y,\\
\log\frac{\mu^{2}}{x^{2}}, & \mbox{if}\; x=y\end{array}\end{cases}~.~\,
\end{eqnarray}
From the above formulae of  $\omega,~\lambda,~\sigma$, the first terms of
Eqs. (\ref{omg-all}), (\ref{sig-all}) and  (\ref{lam-all})
in the brackets come from the quark loops and the rest contributions come from the squark loops.

As a double check, we find that our results can be reduced to the 
previous results for example in Refs.~\cite{Martin:2009bg,Graham:2009gy}, 
if we ignore  the bilinear terms.
Also, the most important feature of this work is that we notice the 
bilinear-term coefficients' contributions to the Feynman rules,
and then to the Higgs boson mass corrections. Thus,
we need to study it and compare the results here.

With the above formulae, we can divide the radiative corrections to the Higgs boson mass into
three cases to study.

\begin{itemize}
\item{\em Case without bilinear terms}
\end{itemize}

In order to study the physics of the bilinear terms, first we should learn what
will happen if we do not have bilinear terms. From superpotential in Eq.~(\ref{sup-sip}) and the
mass matrices in Eqs.~(\ref{fermmass}) and (\ref{sumasss}),
we can see that if we set $M_{y_q}=0$ and $M_{y_u}=0$, $X$-type quarks/squarks
and $Y$-type quarks will form separate sectors.
Their contributions to $\omega,~\lambda,~\sigma$ can be divided into
$\omega_x,~\lambda_x,~\sigma_x$ and $\omega_y,~\lambda_y,~\sigma_y$,  which are given by
\begin{eqnarray}
\omega_x &=&-\frac{N_{c}G_{F}m_{x}^{4}}{\sqrt{2}\pi^{2}\sin^{2}\beta}
\Biggl[\log\frac{m_{\tilde{x}_{1}}m_{\tilde{x}_{2}}}{m_{x}^{2}}
+\frac{A_{x}(A_{x}-\mu\cot\beta)}{m_{\tilde{x}_{1}}^{2}-m_{\tilde{x}_{2}}^{2}}
\log\frac{m_{\tilde{x}_{1}}^{2}}{m_{\tilde{x}_{2}}^{2}}\nonumber\\
&& +\frac{A_{x}^{2}(A_{x}-\mu\cot\beta)^{2}}{(m_{\tilde{x}_{1}}^{2}
-m_{\tilde{x}_{2}}^{2})^{2}}(1-\frac{m_{\tilde{x}_{1}}^{2}
+m_{\tilde{x}_{2}}^{2}}{m_{\tilde{x}_{1}}^{2}-m_{\tilde{x}_{2}}^{2}}
\log\frac{m_{\tilde{x}_{1}}}{m_{\tilde{x}_{2}}})\Biggr]~,~\\
\omega_y &=&-\frac{N_{c}G_{F}m_{y}^{4}}{\sqrt{2}\pi^{2}\cos^{2}\beta}
\left[\frac{\mu^{2}(A_{y}-\mu\tan\beta)^{2}}{(m_{\tilde{y}_{1}}^{2}
-m_{\tilde{y}_{2}}^{2})^{2}}(1-\frac{m_{\tilde{y}_{1}}^{2}
+m_{\tilde{y}_{2}}^{2}}{m_{\tilde{y}_{1}}^{2}
-m_{\tilde{y}_{2}}^{2}}\log\frac{m_{\tilde{y}_{1}}}
{m_{\tilde{y}_{2}}})\right]~,~
\label{equ1}
\end{eqnarray}
\begin{eqnarray}
\lambda_x &=&-\frac{N_{c}G_{F}m_{x}^{4}}{\sqrt{2}\pi^{2}\sin^{2}\beta}
\Biggr[\frac{\mu(A_{x}-\mu\cot\beta)}{m_{\tilde{x}_{1}}^{2}
-m_{\tilde{x}_{2}}^{2}}\log\frac{m_{\tilde{x}_{1}}^{2}}{m_{\tilde{x}_{2}}^{2}}\nonumber\\
&&+\frac{2\mu A_{x}(A_{x}-\mu\cot\beta)^{2}}{(m_{\tilde{x}_{1}}^{2}
-m_{\tilde{x}_{2}}^{2})^{2}}(1-\frac{m_{\tilde{x}_{1}}^{2}
+m_{\tilde{x}_{2}}^{2}}{m_{\tilde{x}_{1}}^{2}-m_{\tilde{x}_{2}}^{2}}
\log\frac{m_{\tilde{x}_{1}}}{m_{\tilde{x}_{2}}})\Biggr]~,~\\
\lambda_y &=&-\frac{N_{c}G_{F}m_{y}^{4}}{\sqrt{2}\pi^{2}\cos^{2}\beta}
\Biggr[\frac{\mu(A_{y}-\mu\tan\beta)}{m_{\tilde{y}_{1}}^{2}
-m_{\tilde{y}_{2}}^{2}}\log\frac{m_{\tilde{y}_{1}}^{2}}{m_{\tilde{y}_{2}}^{2}}\nonumber\\
&&+\frac{2\mu A_{y}(A_{y}-\mu\tan\beta)^{2}}{(m_{\tilde{y}_{1}}^{2}
-m_{\tilde{y}_{2}}^{2})^{2}}(1-\frac{m_{\tilde{y}_{1}}^{2}
+m_{\tilde{y}_{2}}^{2}}{m_{\tilde{y}_{1}}^{2}-m_{\tilde{y}_{2}}^{2}}
\log\frac{m_{\tilde{y}_{1}}}{m_{\tilde{y}_{2}}})\Biggr]~,~
\label{equ2}
\end{eqnarray}
\begin{eqnarray}
\sigma_x &=&-\frac{N_{c}G_{F}m_{x}^{4}}{\sqrt{2}\pi^{2}\sin^{2}\beta}
\left[\frac{\mu^{2}(A_{x}-\mu\cot\beta)^{2}}{(m_{\tilde{x}_{1}}^{2}
-m_{\tilde{x}_{2}}^{2})^{2}}(1-\frac{m_{\tilde{x}_{1}}^{2}
+m_{\tilde{x}_{2}}^{2}}{m_{\tilde{x}_{1}}^{2}-m_{\tilde{x}_{2}}^{2}}
\log\frac{m_{\tilde{x}_{1}}}{m_{\tilde{x}_{2}}})\right]~,~\\
\sigma_y &=&-\frac{N_{c}G_{F}m_{y}^{4}}{\sqrt{2}\pi^{2}\cos^{2}\beta}
\Biggr[\log\frac{m_{\tilde{y}_{1}}m_{\tilde{y}_{2}}}{m_{y}^{2}}
+\frac{A_{y}(A_{y}-\mu\tan\beta)}{m_{\tilde{y}_{1}}^{2}
-m_{\tilde{y}_{2}}^{2}}\log\frac{m_{\tilde{y}_{1}}^{2}}{m_{\tilde{y}_{2}}^{2}}\nonumber\\
&&+\frac{A_{y}^{2}(A_{y}-\mu\tan\beta)^{2}}{(m_{\tilde{y}_{1}}^{2}
-m_{\tilde{y}_{2}}^{2})^{2}}(1-\frac{m_{\tilde{y}_{1}}^{2}
+m_{\tilde{y}_{2}}^{2}}{m_{\tilde{y}_{1}}^{2}-m_{\tilde{y}_{2}}^{2}}
\log\frac{m_{\tilde{y}_{1}}}{m_{\tilde{y}_{2}}})\Biggr]~.~
\label{equ3}
\end{eqnarray}
Note that, in this case the vector-like quark masses are
\begin{eqnarray}
  \label{eq:fma}
  m_x = \frac{Y_x}{\sqrt{2}} v_u, ~~m_y= \frac{Y_y}{\sqrt{2}} v_d~,~\,
\end{eqnarray}
which have been used in the formulae.
\begin{figure}[!htbp]
\centering
    \includegraphics[width=.9\textwidth]{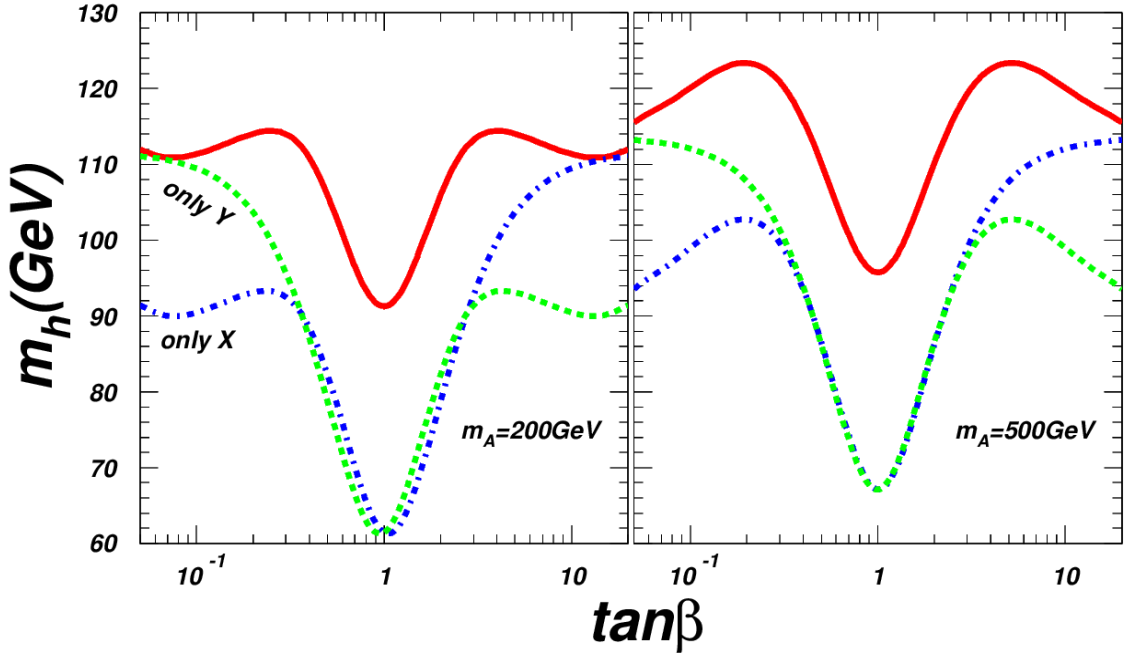}
    \caption{The Higgs boson mass $m_h$ versus $\tan\beta$ with  different $M_A$.
     The solid, dot dash and dash lines denotes the total contribution,  the contributions
     solely from $X$-type VLPs and $Y$-type VLPs, respectively.}
    \label{fig1}
\end{figure}

As we discussed above, $\omega_x,~\lambda_x,~\sigma_x$ are exactly same
as the leading contributions of top quark and squark in the MSSM in form. 
Furthermore, from Eq.~(\ref{equmh1}) and
the superpotential in Eq.~(\ref{sup-sip}), we get that
if setting all the corresponding input parameters the same, 
under $\tan\beta\leftrightarrow \cot\beta$, 
the contributions to the radiative correction from
$X$-type VLPs and $Y$-type VLPs are symmetric.
The point is that if we neglect the Yukawa couplings of the third generation especially top quarks/squarks,
 there is a symmetry between $H_d$ and $H_u$. However, $\tan\beta$ cannot
be smaller than about 2 for consistency in the supersymmetric SMs.

To illustrate the symmetry more clearly, we
plot the corrected Higgs mass $m_h$ versus the $\tan\beta$ with different input $M_A$ in Fig.~\ref{fig1}.
The $\tan\beta$ axis is in logarithmic coordinate.
We can see that both panels are mirror symmetric on the axis $\tan\beta=1$.
In our numerical calculations all other parameters are taken as
$$\mu=200{~\rm GeV},~m_x=m_y=200{~\rm GeV},~m_{\tilde x_1}=m_{\tilde x_1}=450{~\rm GeV},~
m_{\tilde x_2}=m_{\tilde x_2}=550{~\rm GeV},~A_x=A_y=1500{~\rm GeV},~$$
for the left panel and
$$\mu=50{~\rm GeV},~m_x=m_y=200{~\rm GeV},~m_{\tilde x_1}=m_{\tilde x_1}=450{~\rm GeV},~
m_{\tilde x_2}=m_{\tilde x_2}=550{~\rm GeV},~A_x=A_y=1000{~\rm GeV},~$$
for the right panel.

\begin{figure}[!htbp]
\begin{minipage}[t]{0.48\textwidth}
    \centering
    \includegraphics[width=1.08\textwidth]{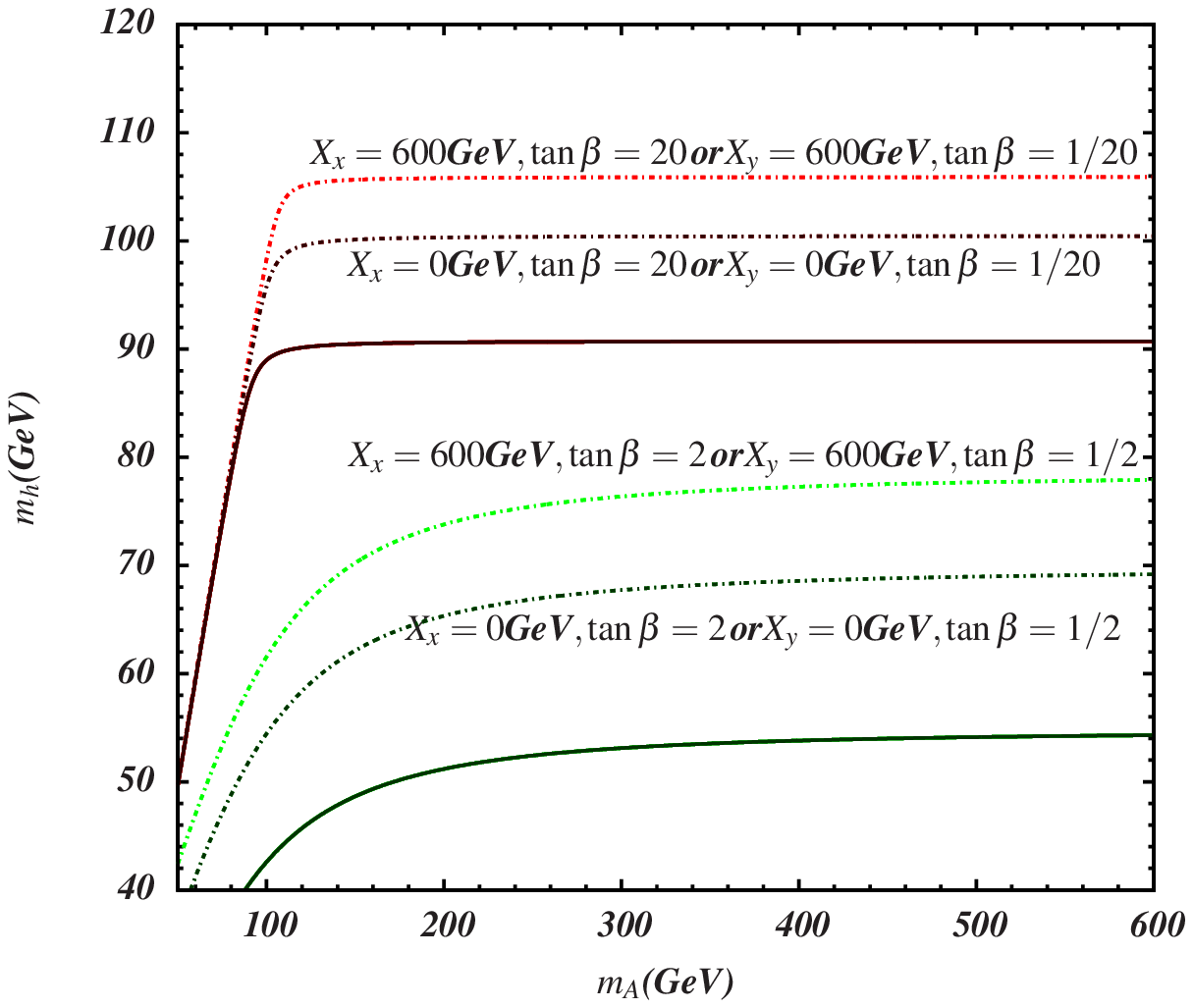}
\end{minipage}
\vspace{0.01\textwidth}
\begin{minipage}[t]{0.48\textwidth}
    \centering
    \includegraphics[width=0.9\textwidth]{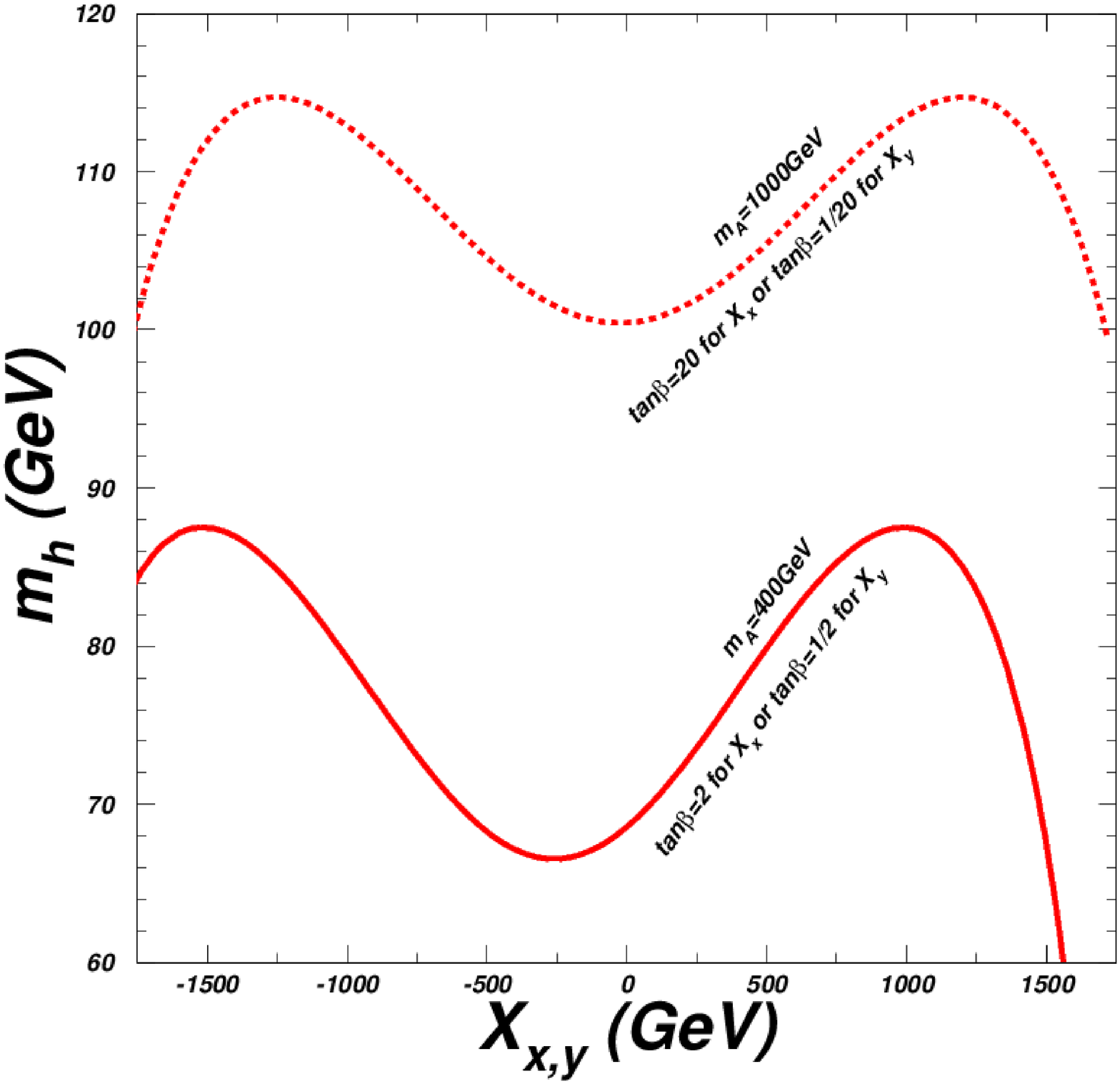}
\end{minipage}
    \caption{The Higgs boson mass $m_h$ versus $m_A$ and $X_x$ or $X_y$, The trend of $X$-type
    quark is the same as the top quark. The result of $Y$-type quark is exactly
    the same as $X$-type quark by changing $\tan\beta$ to $1/\tan\beta$.}
    \label{fig2}
\end{figure}

Following the convention, we define variables
\begin{eqnarray}
  X_x &\equiv & A_{x}-\mu\cot\beta\,,\nonumber\\
  X_y &\equiv & A_{y}-\mu\tan\beta
\end{eqnarray}
for the next step. It is also clear that if we do the transformation:
\begin{eqnarray}
  \label{eq:interch}
  \omega_x \leftrightarrow \sigma_y,~
  \sigma_x \leftrightarrow \omega_y,~
  \lambda_x \leftrightarrow \lambda_y,~
X_x \leftrightarrow X_y,
\end{eqnarray}
the contributions from $X$ and $Y$ particles are symmetric, as displayed in Fig.~\ref{fig2}.

Another very important feature we find is the contribution of $X$-type VLPs is dominant
in case of large $\tan\beta$ while the contribution of $Y$-type VLPs is dominant
in case of large $\cot\beta$. This is conflict with our intuitive judgement, because 
the Yukawa coupling of $X$-type quark is $\cot\beta$ enhanced and $Y$-type quark
is $\tan\beta$ enhanced. However, one can confirm this conclusion after careful checking 
by comparing Eq.~(\ref{equ1}) with Eq.~(\ref{equ3}).
The reason is that the enhanced terms of $\omega,~\sigma,~\lambda$ canceled each other,
leaving an enhancement trend that is beyond simple intuition. 
For example, in case of large $\cot\beta$, the
enhanced $\omega_x,~\sigma_x$ in first term of Eq.~(\ref{equmh1}) are cancelled by the
second root term, making the radiative correction
dominantly from $\omega_y,~\lambda_y,~\sigma_y$.

Because the radiative corrections are symmetric on the interchange of
$\tan\beta$ and $\cot\beta$ if we neglect the third generation, we can image that 
the sole corrections from
the $X$-type quarks and those from $Y$-type quarks will be exactly
the same by changing $\tan\beta$ to its inverse. This is clearly shown in Fig.~\ref{fig2}.
Fig.~\ref{fig2} shows $m_h$ versus $m_A$ and $X_x$ or $X_y$ individually.
In the left panel, we set $X_x=0~{\rm and}~600~\rm GeV$, $\tan\beta=20~{\rm and}~2$,
 and vary $M_A$,
while in the right panel, we vary $X_x$ or $X_y$ at different $M_A$.
From the left panel, we can also see that when $M_A$ increases, 
the heavy Higgs particles will decouple at the EW scale and then there is one SM like Higgs
with mass about 100 GeV, but the SUSY radiative corrections can lift the Higgs boson mass.
The right panel shows the effects of the mixings among
squarks. Because $X_x$ and $X_y$ determine the mixings between $\tilde x_L,~\tilde x_R$ and
$\tilde y_L,~\tilde y_R$, there is no mixing in squark sector when $X_{x,y}=0$, and then the radiative
corrections only depend on the mass splittings between quarks and squarks, namely
the first terms of $\omega_x$ and $\sigma_y$. An appropriate $X_x$ or
$X_y$ can give large logarithmic terms $\log({m^2_{\tilde{x}_1}}/{m^2_{\tilde{x}_2}})$ or
$\log({m^2_{\tilde{y}_1}}/{m^2_{\tilde{y}_2}})$, which will
enhance the radiative corrections.

\begin{figure}[!htbp]
    \centering
    \includegraphics[width=0.5\textwidth]{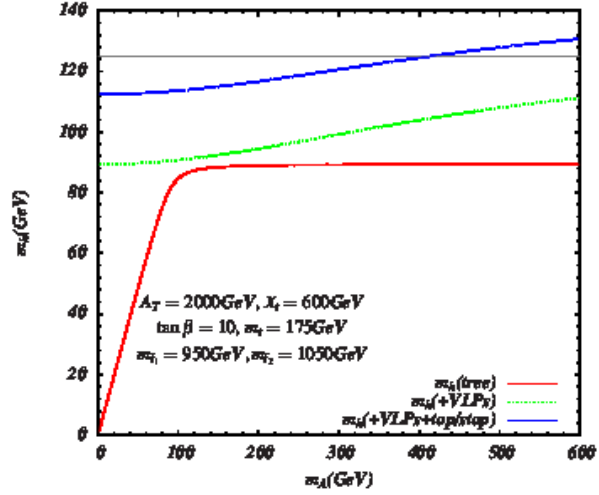}
    \caption{The Higgs boson mass $m_h$ versus $m_A$ to reflect the mass.
             The numerical value of the gray line is $125$~GeV.}
    \label{fig22}
\end{figure}

 To calculate the Higgs boson mass more precisely, 
we add the top quark/squark contributions, and the analytical 
one-loop contributions to the Higgs boson from the top quarks/squarks 
are
\begin{eqnarray}
&&\omega_{t}=\omega_{x}(m_{x}\to m_{t},m_{\tilde{x}_{1}}\to m_{\tilde{t}_{1}},m_{\tilde{x}_{2}}\to m_{\tilde{t}_{2}},A_{x}\to A_{t}),\\
&&\lambda_{t}=\lambda_{x}(m_{x}\to m_{t},m_{\tilde{x}_{1}}\to m_{\tilde{t}_{1}},m_{\tilde{x}_{2}}\to m_{\tilde{t}_{2}},A_{x}\to A_{t}),\\
&&\sigma_{t}=\sigma_{x}(m_{x}\to m_{t},m_{\tilde{x}_{1}}\to m_{\tilde{t}_{1}},m_{\tilde{x}_{2}}\to m_{\tilde{t}_{2}},A_{x}\to A_{t}).
\end{eqnarray}
Thus the complete $\omega,\,\lambda,\,\sigma$ should be 
\begin{eqnarray}
&&\omega=\omega_{x}+\omega_{y}+\omega_{t},\\
&&\lambda=\lambda_{x}+\lambda_{y}+\lambda_{t},\\
&&\sigma=\sigma_{x}+\sigma_{y}+\sigma_{t},
\end{eqnarray}
and we choose the same input parameters as in Fig.~\ref{fig2}, except $\tan\beta=10$. 
For $A_T=2000$~GeV, $m_{\tilde t_1}=950$~GeV, and $m_{\tilde t_2}=1050$~GeV,
we present the Higgs boson mass $m_h$ versus $m_A$ in Fig.~\ref{fig22}.

In Fig.~\ref{fig22}, the red solid line represents the tree-level Higgs mass, 
the green dashed line represents the Higgs mass with the VLPs' contributions,
and the blue dashed line represents the Higgs mass with both the VLPs' contributions 
and the third generation's contributions. Clearly, the corrections from VLPs 
is positive, but not enough for a 125~GeV Higgs. After including the corrections from the third generation, 
 realizing a $125$~GeV Higgs boson is not difficult in this parameter space.

\begin{itemize}
\item {\em Case with the (partial) decoupling limit}
\end{itemize}

In a supersymmetric theory, the masses of fermion(s) and sfermion(s) 
in a chiral superfield are degenerated.
The soft breaking terms are added to the Lagrangian
to split the spectra of the fermions and sfermions in the phenomenological models.
Thus, there is a limit with vanishing super-trace, {\it i.e.}, ${\rm Str} M^2=0$. 
We find this limit would be a
very special situation to the MSSMV, 
since the bilinear term coefficients will be more prominent.

%{\red In the MSSM with exact supersymmetry, the Higgs VEVs are zero. So I do not get
%the point why we can choose non-zero VEVs. The way to get the VEVs is that we introduce
%a SM singlet $S$, and consider the superpotential $S(H_dH_u-v^2)$. }{\blue I have no idea.}

If soft breaking parameters in the quark/squark sector are all ignored,
we only have the supersymmetric lagrangian generated from the superpotential in
Eq.~(\ref{sup-sip}) for quarks and squarks. For example, the
mass matrix of quarks is
\begin{eqnarray}
\left(\begin{array}{cc}
\frac{1}{\sqrt{2}}Y_{x}v_{u} & M_{y_q}\\
M_{y_u} & \frac{1}{\sqrt{2}}Y_{y}v_{d}
\end{array}\right).\label{vfer}
\end{eqnarray}
So the $X$-type quarks and $Y$-type quarks can be mixed.
The mass matrix for squarks (dropping all the soft term of Eq.~(\ref{sumasss})) is
\begin{eqnarray}
%\hspace{-10mm}
\left(\begin{array}{cccc}
(\frac{v_{u}}{\sqrt{2}}Y_{x})^{2}+M_{y_q}^{2}
& 0 & 0 & M_{y_u}\frac{v_u}{\sqrt{2}}Y_{x}+M_{y_q}\frac{v_{d}}{\sqrt{2}}Y_{y}\\
0 & (\frac{v_{u}}{\sqrt{2}}Y_{x})^{2}+M_{y_u}^{2}
& M_{y_q}\frac{v_{u}}{\sqrt{2}}Y_{x}+M_{y_u}\frac{v_d}{\sqrt{2}}Y_{y} & 0\\
0 & M_{y_q}\frac{v_u}{\sqrt{2}}Y_{x}+M_{y_u}\frac{v_d}{\sqrt{2}}Y_{y}
& (\frac{v_d}{\sqrt{2}}Y_{y})^{2}+M_{y_q}^{2}
  & 0\\
M_{y_u}\frac{v_u}{\sqrt{2}}Y_{x}+M_{y_q}\frac{v_d}{\sqrt{2}}Y_{y}
& 0 &  0
& (\frac{v_d}{\sqrt{2}}Y_{y})^{2}+M_{y_u}^{2}
\end{array}\right).
\end{eqnarray}
Although there are four squarks in the spectra, they can be divided into
two pairs with degenerated masses. Futhermore, one can check that these two masses
 are also degenerated with the masses of quarks listed Eq.~(\ref{vfer}), which
is just required by a supersymmetric theory because
the four squarks are the super partners of the two quarks.

\begin{figure}[!htbp]
    \centering
    \includegraphics[width=.45\textwidth]{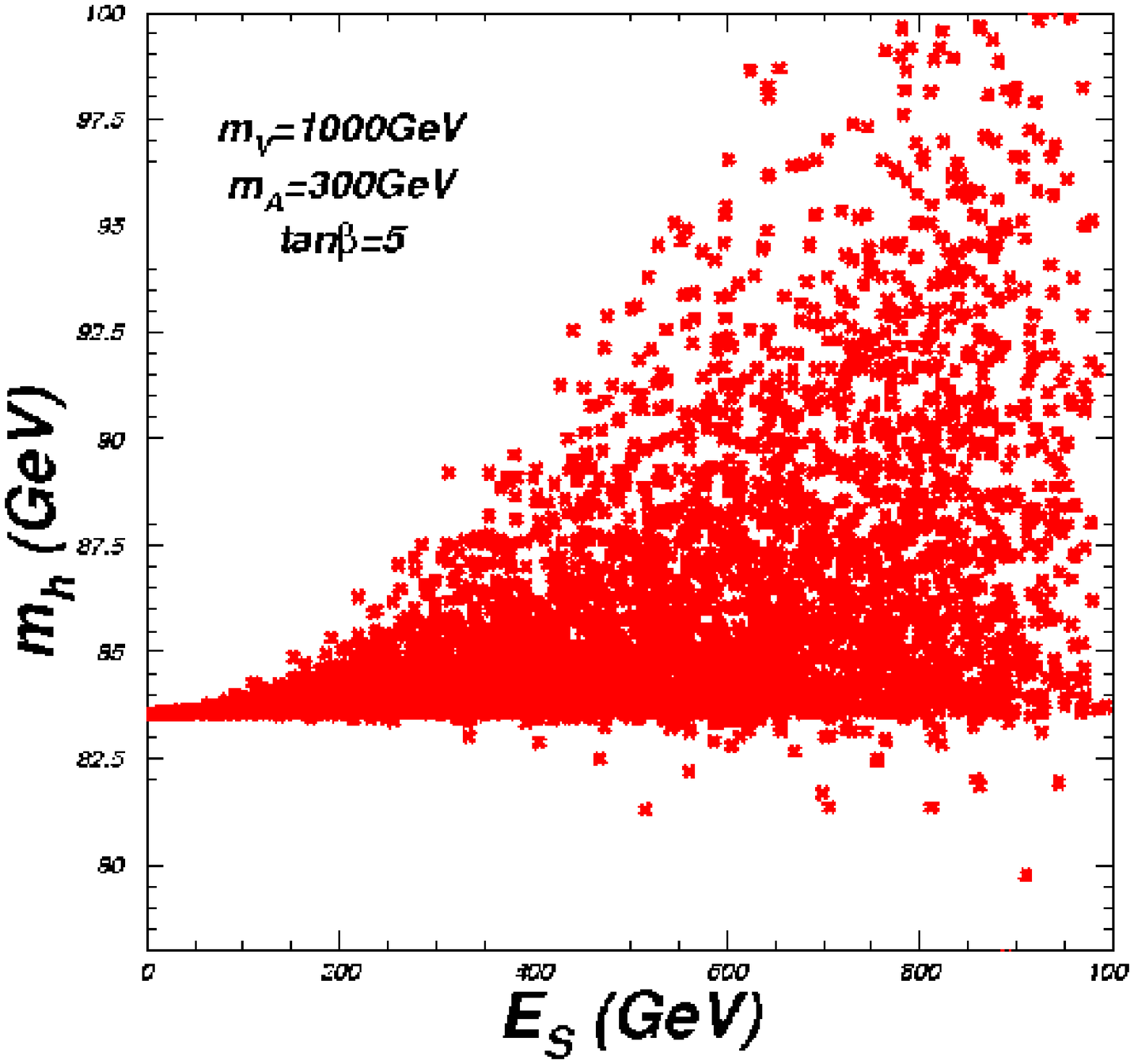}
    \caption{The Higgs mass $m_h$ versus EW energy scale $E_S$ in case of
    $M_A =300~\rm GeV$, $\tan\beta=5$ and $M_{y_q}=M_{y_u}=M_V=1000~\rm GeV$.  All the other
    soft parameters are scaned randomly in the range $(-E_S,~E_S)$. The tree-level
    SM-like Higgs mass is $83.56~\rm GeV$.}
    \label{fig3}
\end{figure}

To test this supersymmetic limit, we choose $g_1=g_2 = 0$ to remove gauge interactions. 
We take $M_{y_q}=M_{y_u}=M_V=1000~\rm GeV$, $M_A =300~\rm GeV$, $\tan\beta=5$,
and the mass of tree-level Higgs is $83.56~\rm GeV$. And then we scan
the  Yukawa couplings $Y_x,~Y_y$ randomly in the range of $(0,~~2)$ and
the other soft paramters (together with $\mu$)
$$A_x,~A_y,~\sqrt{B_{M_{y_q}} M_{y_q}},~\sqrt{B_{M_{y_u}} M_{y_u}},$$
randomly in the range of $(-E_S, +E_S)$ with the parameter $E_S$ denoting the EW
energy scale and varing randomly in the range of $(0{~\rm GeV},~~1000{~\rm GeV})$.
The  radiative masses of the SM-like Higgs boson
versus the EW energy scale are shown in Fig.~\ref{fig3}.
In Fig. \ref{fig3},
we can see an interesting feature of the MSSMV that the radiative
corrections decrease to zero as $E_S$ decreases to zero, and this trend is 
independent of the billinear parameters $M_{y_q}$ and $M_{y_u}$. 
The reason is that fermions and sfermions cancel each other in a 
supersymmetric theory.

%{\red With non-zero supersymmetry breaking soft paramters, there is NO supersymmetric limit. 
%The limit, I think, is the simplified limit without gauge interactions. } {\blue I have no idea.}

\begin{figure}[!htbp]
    \centering
    \includegraphics[width=.8\textwidth]{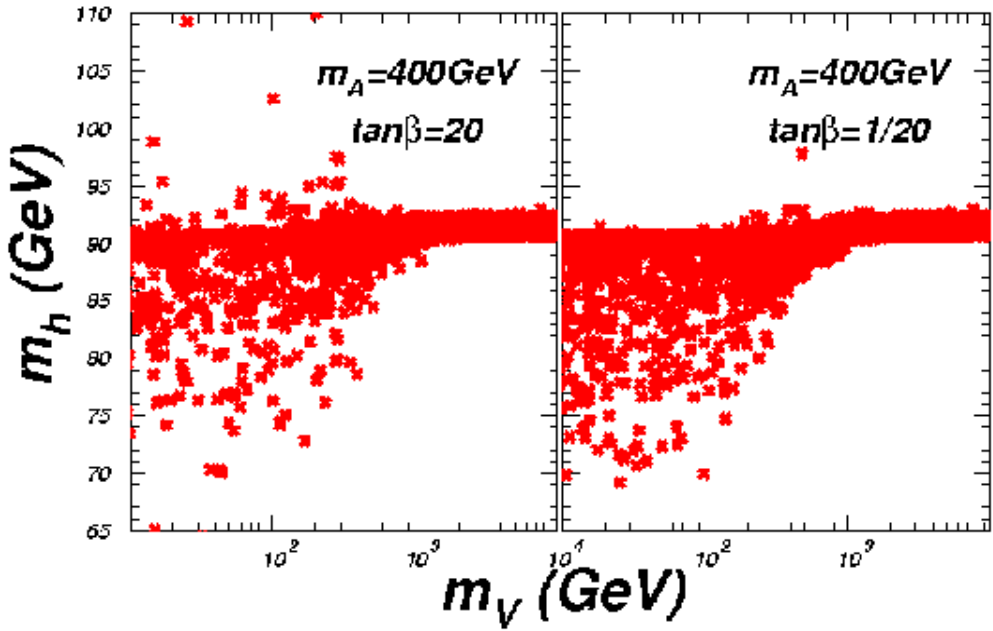}
    \caption{The Higgs mass $m_h$ versus VLP input $M_{y_q}=M_{y_u}=M_V$,
    in case of $M_A =400~\rm GeV$, $\tan\beta=20,1/20$ and $E_S=1000~\rm GeV$ .
    All the other soft parameters are scaned randomly in the range $(-E_S,~E_S)$.
    The tree-level SM-like Higgs mass is $83.56~\rm GeV$.}
    \label{fig4}
\end{figure}

Now let us pay more attention to understand the decoupling limit
in the MSSMV.  If $M_{y_q},~M_{y_u}\gg E_S$, the model will recover supersymmetry,
and then the effects of VLPs will decouple from the EW energy scale, namely,
the radiative corrections to Higgs boson mass will be zero, 
as demonstrated in Fig.~\ref{fig4}.

In Fig.~\ref{fig4}, we choose $E_S =1000~\rm GeV$, $M_A =400~\rm GeV$, and scan $M_{y_q}=M_{y_u}=M_V$ randomly
from $0~\rm GeV$  to $10~\rm TeV$, $\mu $ and all the soft paramters in squark sector
randomly in the range of $(-E_S,~E_S)$ with $g_1$ and $g_2$ from the SM inputs.
Note that the symmetry between $\tan\beta\leftrightarrow \cot\beta$ still remains.
We set $\tan\beta =20$ (left panel) and $\tan\beta =1/20$ (right panel)
for demonstration. From Fig.~\ref{fig4}, we can see that the radiative corrections to
 the Higgs boson mass become a very small value as the $M_V$ increases up to be much heavier
than $1~\rm TeV$. Thus, the VLPs decouple from the EW energy scale.
The decoupling effects can also be seen in Refs. \cite{Martin:2009bg,Huo:2011zt}.
As the masses of VLPs increase, the splittings between the quarks and squarks as well as the splitings
among squarks become more and more negligible. 
Namely, the term
\begin{eqnarray}
  \log \frac{M_S^2+M_V^2}{M_V^2}
\end{eqnarray}
approaches to zero, and then the VLPs decouple.

From the above two cases, we conclude that
\begin{enumerate}
\item The corrections to the Higgs boson mass in the MSSMV depend on the splitting
 between quarks and squarks 
and the splitting among squarks. The soft terms break supersymmetry
explicitly, leading to the splitting of the mass spectra.
\item The heavy VLPs can suppress the splitting of the mass spectra.
Because there exists the decoupling limit, the heavy VLP effects 
decouple from the EW energy scale.
\end{enumerate}

\begin{itemize}
\item {\em A partial decoupling effect}
\end{itemize}

Now we turn to the case that one VLP input $M_{y_q}$ is at the EW energy scale, but the other one 
$M_{y_u}$ is free. We find that there exists the decoupling effect as well.
Let us take
\begin{equation}
  \label{eq:mmm}
  M_{y_u}\gg M_{y_q}~~~, 
\end{equation}
easily we would get one light VLP quark and one heagy VLP quark.
So there exist two light squarks and two heavy squarks, and
there can be large splittings between quarks and squarks. 
However, the effects of these large splittings can be suppressed
by the heavy particles, the corrections from VLPs including the light and heavy
ones decouple at the EW energy scale.
\begin{figure}[!htbp]
    \centering
    \includegraphics[width=.45\textwidth]{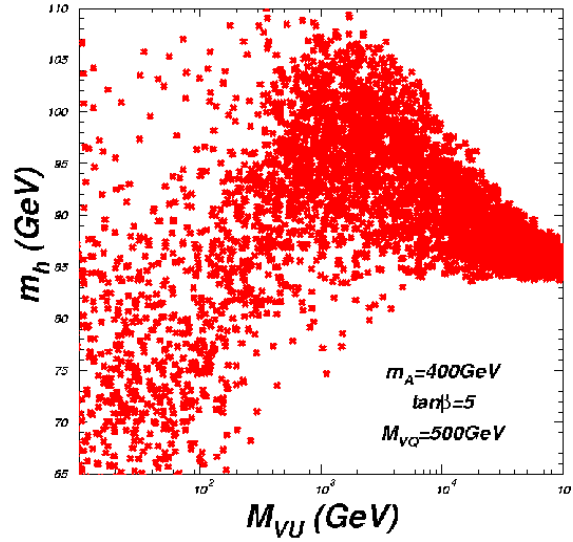}
    \caption{The radiative Higgs boson mass $m_h$ versus VLP input $M_{y_u}$
    in case of $M_A =400~\rm GeV$, $\tan\beta=5$,
    $M_{y_q}=500~\rm GeV$ and $E_S=1000~\rm GeV$.
    All the other soft parameters are scaned randomly in the range $(-E_S,~E_S)$.
    The tree-level SM-like Higgs mass is $83.83~\rm GeV$.}
    \label{fig5}
\end{figure}

To show  this effect,
we choose $M_A=400~\rm GeV$, $\tan\beta=5$ and $M_{y_q}=500~\rm GeV$, and
scan $M_{y_u}$ in the range of $(0{~\rm GeV},~~100{~\rm TeV})$.
$\mu $ and all the soft parameters in squark sector vary
randomly in the range of $(-1000{~\rm GeV},~1000{~\rm GeV})$. The radiative
corrections to the Higgs boson mass are given in Fig.~\ref{fig5}. We can
see that as $M_{y_u}$ increases (bigger than $M_{y_q}$), 
the virtual effects of VLPs on Higgs mass become smaller and smaller. 
Finally, the VLP effects decouple from the EW energy scale, 
though there are still light quarks and squarks that are
around the EW energy scale.

Such suppression can be understood in the similar way as
the decoupling limit mentioned above.
The splitting between quarks and squarks, and splitting between squarks are still
suppressed by the heavy VLPs, so the partial decoupling limits still exist. 
In fact, after we integrate out $\hat X_q$ and $\hat Y_q$, we get
\begin{eqnarray}
W&\supset & \frac{Y_{x_u}Y_{y_u} \hat X_u \hat Y_u \hat{H}_{u} \hat{H}_{d}}{M_{y_q}}~,~
\end{eqnarray}
which will be decoupled for very heavy $M_{y_q}$.

\begin{figure}[htbp]
    \centering
    \includegraphics[width=.8\textwidth]{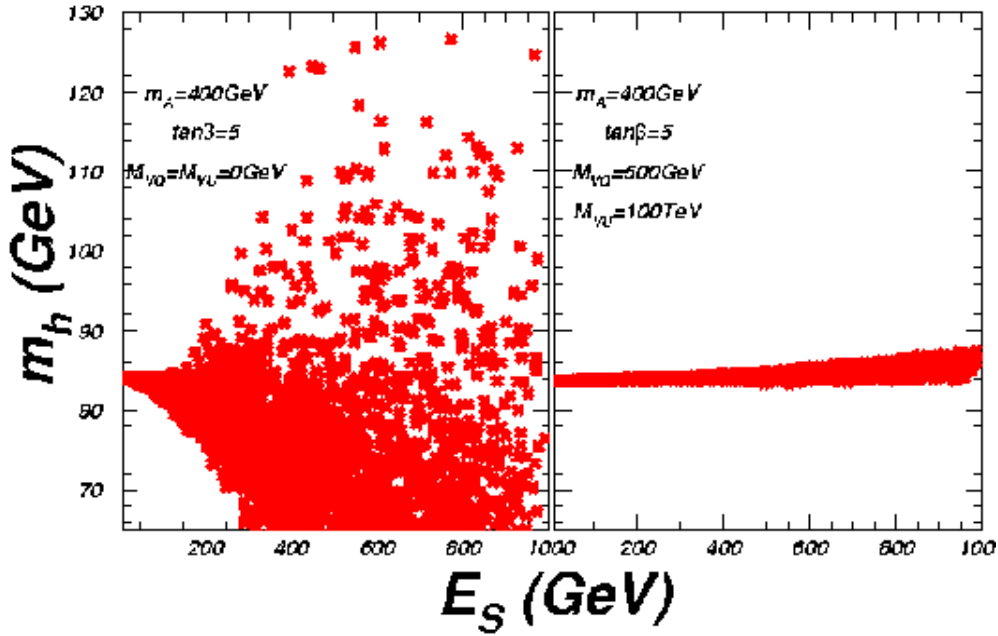}
    \caption{The Higgs mass $m_h$ versus the EW energy scale  $E_S$
    in case of  $M_A =400\rm GeV$, $\tan\beta=5$, $M_{y_q}=500\rm GeV$, and $M_{y_u}=100\rm TeV$.
    All the other soft parameters are scanned randomly in the range $(-E_S,~E_S)$.
   The tree-level SM-like Higgs mass is $83.83~\rm GeV$.}
    \label{fig6}
\end{figure}

At the end of this Section, we compare the Higgs mass in the cases {\em with} and {\em without}
the bilinear terms.
The numerical results are shown in Fig.~\ref{fig6}. The left panel shows the results
without the bilinear terms (corresponding to the first case),
while right panel shows the results
with bilinear terms $M_{y_q}=500{~\rm GeV},~M_{y_u}=100~\rm TeV$. 
We can see that the mass of Higgs boson mass can be enhanced in the first case. 
However, the virtual effects of VLPs are much smaller in case of heavy partners existing.
Two benchmark points are presented in Table \ref{tab2}.
\begin{table}[h]
\centering
\begin{tabular}{cccccccc}
\hline
\hline
 (GeV) &$m_{q_1}(m_{x})$ & $m_{q_2}(m_{y})$  &$m_{\tilde q_1}(m_{\tilde x_1})$
 &  $m_{\tilde q_2}(m_{\tilde x_2})$
 & $m_{\tilde q_3}(m_{\tilde y_1})$ & $m_{\tilde q_4}(m_{\tilde y_2})$  & $m_h$\\
\hline
Without billinear terms & 48 &  240 &  102 &  373 & 444 & 690 & 126.2\\
\hline
One heavy billinear terms  & 500 &  100000 &  153 &  827 & 100001 & 100003 & 84.2\\
\hline
\hline
\end{tabular}
\caption{The benchmark points of particle spectra with and without heavy VLPs for $\tan\beta=5$.
   The tree-level SM-like Higgs mass is $83.83~\rm GeV$. }
\label{tab2}
\end{table}
Note that $X$-type particle just like the top quark except its bilinear terms for mass source, 
but if there is a heavy VLP parter of this particle, 
their corrections to Higgs mass will be negligible. 
The reason for this can also be understood from the Feynman rules
of $hq_i q_j$ and $h\tilde q_i \tilde q_j$. The couplings between Higgs and two light
quarks are also suppressed by the rotation matrices, leading the radiative
results of loops are negligible. On the other hand, if we want the radiative corrections from the
$X$-type VLP quarks and squarks to enhance the tree-level Higgs mass to 125 GeV, 
the $Y$-type VLP partner 
will be strictly constrained. The detail study of such phenomenology is beyond this work.

\section{Summary}
\label{sec:summary}
We proposed the MSSMV, and studied the particle spectra
and Feynman rules in the quark/lepton and squark/slepton sectors.
Due to the participation of bilinear terms, we found three different points 
compared to the Feynman rules in the MSSM
\begin{enumerate}
\item The tree-level FCNC processes in the quark/lepton sectors.
\item Deviation from unitarity.
\item Interactions from F-term potential of the bilinear terms.
\end{enumerate}
Using these Feynman rules, we studied the analytical one-loop radiative contributions to Higgs mass
with one pair of VLPs. We found a very interesting mechanism in the MSSMV.
All the effects of VLPs can decouple from the EW energy scale if the bilinear terms
are much heavy. And if there are one light bilinear term and one heavy
bilinear term, the virtual effects of the light one can be suppressed by the heavy one.
This suppression can make the EW phenomenology highly different from the MSSM.
Note that in our above study of the radiative corrections to Higgs mass,
we ignored the experimental limits on the input parameters such as $\tan\beta$,
$M_S$, etc. A complete study will be given elsewhere.

%%%%%%%%%%%%%%%%%%%%%%%%%%%%%%%%%%%%%%%%%%%%%%%
\section*{Acknowledgment}
\label{sec:ack}

The work of T.L. is supported by 
the Natural Science Foundation of China under grant numbers 11135003, 11275246, and 11475238.
The work of W.~Y. Wang was supported by the Natural Science Foundation of China under grant numbers 11375001, 
the Ri-Xin Foundation of BJUT and Youth-Talents Foundation of eduction department of Beijing.

\section*{Appendix A}
\label{sec:app1}
~\\
We shall list the Feynman rules for the interactions
involving VLPs, where all the interaction state summing subscripts run from 1 to 4.

\subsection*{1. Fermion-Higgs Boson}
%% fermions h
~\\
%%%%%%%%%%%%%%%%%%%%%%%%%%%%%%%%%%%%
\begin{tabular}{ll}
\fcolorbox{white}{white}{
  \begin{picture}(157,104) (19,5)
    \SetWidth{1.0}
    \SetColor{Black}
    \Line[arrow,arrowpos=0.5,arrowlength=5,arrowwidth=2,arrowinset=0.2](96,72)(96,8)
    \Line[arrow,arrowpos=0.5,arrowlength=5,arrowwidth=2,arrowinset=0.2](96,8)(32,8)
    \Line[dash,dashsize=5,arrow,arrowpos=0.5,arrowlength=5,arrowwidth=2,arrowinset=0.2](96,8)(160,8)
    \Text(16,4)[lb]{\Large{\Black{$\bar{d}_{i}$}}}
    \Text(94,74)[lb]{\Large{\Black{$d_{j}$}}}
    \Text(166,4)[lb]{\Large{\Black{$h_{k}$}}}
  \end{picture}
}
&
\raisebox{35\unitlength}{
\begin{minipage}{5cm}
\begin{eqnarray}
&&
-i\frac{1}{\sqrt{2}}\Big[\big(
U_{R,{ia}}^{d,*}Y_{d,{ab}}U_{L,{jb}}^{d,*}Z_{{k1}}^{H}
+Y_{yd}U_{L,{j5}}^{d,*}U_{R,{i5}}^{d,*}Z_{{k2}}^{H}\big)P_L\nonumber\\
&&
+\big(U_{R,{ja}}^{d}Y_{d,{ab}}^{*}U_{L,{ib}}^{d}Z_{{k1}}^{H}
+Y_{yd}U_{L,{i5}}^{d}U_{R,{j5}}^{d}Z_{{k2}}^{H}\big)P_R\Big]\nonumber
\end{eqnarray}
\end{minipage}
}
\end{tabular}

~\\
%%%%%%%%%%%%%%%%%%%%%%%%%%%%%%%%%%%%
\begin{tabular}{ll}
\fcolorbox{white}{white}{
  \begin{picture}(157,104) (19,5)
    \SetWidth{1.0}
    \SetColor{Black}
    \Line[arrow,arrowpos=0.5,arrowlength=5,arrowwidth=2,arrowinset=0.2](96,72)(96,8)
    \Line[arrow,arrowpos=0.5,arrowlength=5,arrowwidth=2,arrowinset=0.2](96,8)(32,8)
    \Line[dash,dashsize=5,arrow,arrowpos=0.5,arrowlength=5,arrowwidth=2,arrowinset=0.2](96,8)(160,8)
    \Text(16,4)[lb]{\Large{\Black{$\bar{u}_{i}$}}}
    \Text(94,74)[lb]{\Large{\Black{$u_{j}$}}}
    \Text(166,4)[lb]{\Large{\Black{$h_{k}$}}}
  \end{picture}
}
&
\raisebox{35\unitlength}{
\begin{minipage}{5cm}
\begin{eqnarray}
&&
-i\frac{1}{\sqrt{2}}\Big[\big(U_{R,{ia}}^{u,*}Y_{u,{ab}}U_{L,{jb}}^{u,*}Z_{{k2}}^{H}
+Y_{yu}U_{L,{j5}}^{u,*}U_{R,{i5}}^{u,*}Z_{{k1}}^{H}\big)P_L\nonumber\\
&&
+\big(U_{R,{ja}}^{u}Y_{u,{ab}}^{*}U_{L,{ib}}^{u}Z_{{k2}}^{H}
+Y_{yu}Z_{{k1}}^{H}U_{L,{i5}}^{u}U_{R,{j5}}^{u}\big)P_R\Big]\nonumber
\end{eqnarray}
\end{minipage}
}
\end{tabular}

~\\
%%%%%%%%%%%%%%%%%%%%%%%%%%%%%%%%%%%%
\begin{tabular}{ll}
\fcolorbox{white}{white}{
  \begin{picture}(157,104) (19,5)
    \SetWidth{1.0}
    \SetColor{Black}
    \Line[arrow,arrowpos=0.5,arrowlength=5,arrowwidth=2,arrowinset=0.2](96,72)(96,8)
    \Line[arrow,arrowpos=0.5,arrowlength=5,arrowwidth=2,arrowinset=0.2](96,8)(32,8)
    \Line[dash,dashsize=5,arrow,arrowpos=0.5,arrowlength=5,arrowwidth=2,arrowinset=0.2](96,8)(160,8)
    \Text(16,4)[lb]{\Large{\Black{$\bar{e}_{i}$}}}
    \Text(94,74)[lb]{\Large{\Black{$e_{j}$}}}
    \Text(166,4)[lb]{\Large{\Black{$h_{k}$}}}
  \end{picture}
}
&
\raisebox{35\unitlength}{
\begin{minipage}{5cm}
\begin{eqnarray}
&&
-i\frac{1}{\sqrt{2}}\Big[\big(U_{R,{ia}}^{e,*}Y_{e,{ab}}U_{L,{jb}}^{e,*}Z_{{k1}}^{H}
+Y_{ye}U_{L,{j5}}^{e,*}U_{R,{i5}}^{e,*}Z_{{k2}}^{H}\big)P_L\nonumber\\
&&
+\big(U_{R,{ja}}^{e}Y_{e,{ab}}^{*}U_{L,{ib}}^{e}Z_{{k1}}^{H}
+Y_{ye}U_{L,{i5}}^{e}U_{R,{j5}}^{e}Z_{{k2}}^{H}\big)P_R\Big]\nonumber
\end{eqnarray}
\end{minipage}
}
\end{tabular}

\subsection*{2. Fermion-Gauge Boson}
%% fermions gamma
~\\
%%%%%%%%%%%%%%%%%%%%%%%%%%%%%%%%%%%%
\begin{tabular}{ll}
\fcolorbox{white}{white}{
  \begin{picture}(157,104) (19,5)
    \SetWidth{1.0}
    \SetColor{Black}
    \Line[arrow,arrowpos=0.5,arrowlength=5,arrowwidth=2,arrowinset=0.2](96,72)(96,8)
    \Line[arrow,arrowpos=0.5,arrowlength=5,arrowwidth=2,arrowinset=0.2](96,8)(32,8)
    \Photon(96,8)(160,8){3}{6}
    \Text(16,4)[lb]{\Large{\Black{$\bar{d}_{i}$}}}
    \Text(94,74)[lb]{\Large{\Black{$d_{j}$}}}
    \Text(166,4)[lb]{\Large{\Black{$\gamma_{\mu}$}}}
  \end{picture}
}
&
\raisebox{35\unitlength}{
\begin{minipage}{5cm}
\begin{eqnarray}
-i\frac{1}{3}e \gamma_{\mu}\nonumber
\end{eqnarray}
\end{minipage}
}
\end{tabular}

~\\
%%%%%%%%%%%%%%%%%%%%%%%%%%%%%%%%%%%%
\begin{tabular}{ll}
\fcolorbox{white}{white}{
  \begin{picture}(157,104) (19,5)
    \SetWidth{1.0}
    \SetColor{Black}
    \Line[arrow,arrowpos=0.5,arrowlength=5,arrowwidth=2,arrowinset=0.2](96,72)(96,8)
    \Line[arrow,arrowpos=0.5,arrowlength=5,arrowwidth=2,arrowinset=0.2](96,8)(32,8)
    \Photon(96,8)(160,8){3}{6}
    \Text(16,4)[lb]{\Large{\Black{$\bar{u}_{i}$}}}
    \Text(94,74)[lb]{\Large{\Black{$u_{j}$}}}
    \Text(166,4)[lb]{\Large{\Black{$\gamma_{\mu}$}}}
  \end{picture}
}
&
\raisebox{35\unitlength}{
\begin{minipage}{5cm}
\begin{eqnarray}
&&
i\frac{2}{3}e \gamma_{\mu}\nonumber
\end{eqnarray}
\end{minipage}
}
\end{tabular}

~\\
%%%%%%%%%%%%%%%%%%%%%%%%%%%%%%%%%%%%
\begin{tabular}{ll}
\fcolorbox{white}{white}{
  \begin{picture}(157,104) (19,5)
    \SetWidth{1.0}
    \SetColor{Black}
    \Line[arrow,arrowpos=0.5,arrowlength=5,arrowwidth=2,arrowinset=0.2](96,72)(96,8)
    \Line[arrow,arrowpos=0.5,arrowlength=5,arrowwidth=2,arrowinset=0.2](96,8)(32,8)
    \Photon(96,8)(160,8){3}{6}
    \Text(16,4)[lb]{\Large{\Black{$\bar{e}_{i}$}}}
    \Text(94,74)[lb]{\Large{\Black{$e_{j}$}}}
    \Text(166,4)[lb]{\Large{\Black{$\gamma_{\mu}$}}}
  \end{picture}
}
&
\raisebox{35\unitlength}{
\begin{minipage}{5cm}
\begin{eqnarray}
&&
-i e \gamma_{\mu}\nonumber
\end{eqnarray}
\end{minipage}
}
\end{tabular}

%% fermions Z
~\\
%%%%%%%%%%%%%%%%%%%%%%%%%%%%%%%%%%%%
\begin{tabular}{ll}
\fcolorbox{white}{white}{
  \begin{picture}(157,104) (19,5)
    \SetWidth{1.0}
    \SetColor{Black}
    \Line[arrow,arrowpos=0.5,arrowlength=5,arrowwidth=2,arrowinset=0.2](96,72)(96,8)
    \Line[arrow,arrowpos=0.5,arrowlength=5,arrowwidth=2,arrowinset=0.2](96,8)(32,8)
    \Photon(96,8)(160,8){3}{6}
    \Text(16,4)[lb]{\Large{\Black{$\bar{d}_{i}$}}}
    \Text(94,74)[lb]{\Large{\Black{$d_{j}$}}}
    \Text(166,4)[lb]{\Large{\Black{$Z_{\mu}$}}}
  \end{picture}
}
&
\raisebox{35\unitlength}{
\begin{minipage}{5cm}
\begin{eqnarray}
&&
\frac{ie}{\sin2\Theta_{W}}\gamma_{\mu}\Big[-\frac{2}{3}\sin^{2}\Theta_{W}\delta_{ij}
+U_{L,{ja}}^{d,*}U_{L,{ia}}^{d} P_L +U_{R,{i5}}^{d,*}U_{R,{j5}}^{d} P_R \Big]\nonumber
\end{eqnarray}
\end{minipage}
}
\end{tabular}

~\\
%%%%%%%%%%%%%%%%%%%%%%%%%%%%%%%%%%%%
\begin{tabular}{ll}
\fcolorbox{white}{white}{
  \begin{picture}(157,104) (19,5)
    \SetWidth{1.0}
    \SetColor{Black}
    \Line[arrow,arrowpos=0.5,arrowlength=5,arrowwidth=2,arrowinset=0.2](96,72)(96,8)
    \Line[arrow,arrowpos=0.5,arrowlength=5,arrowwidth=2,arrowinset=0.2](96,8)(32,8)
    \Photon(96,8)(160,8){3}{6}
    \Text(16,4)[lb]{\Large{\Black{$\bar{u}_{i}$}}}
    \Text(94,74)[lb]{\Large{\Black{$u_{j}$}}}
    \Text(166,4)[lb]{\Large{\Black{$Z_{\mu}$}}}
  \end{picture}
}
&
\raisebox{35\unitlength}{
\begin{minipage}{5cm}
\begin{eqnarray}
&&
\frac{ie}{\sin2\Theta_{W}}\gamma_{\mu}\Big[\frac{4}{3}\sin^{2}\Theta_{W}\delta_{ij}
-U_{L,{ja}}^{u,*}U_{L,{ia}}^{u}P_L-U_{R,{i5}}^{u,*}U_{R,{j5}}^{u}P_R\Big]\nonumber
\end{eqnarray}
\end{minipage}
}
\end{tabular}

~\\
%%%%%%%%%%%%%%%%%%%%%%%%%%%%%%%%%%%%
\begin{tabular}{ll}
\fcolorbox{white}{white}{
  \begin{picture}(157,104) (19,5)
    \SetWidth{1.0}
    \SetColor{Black}
    \Line[arrow,arrowpos=0.5,arrowlength=5,arrowwidth=2,arrowinset=0.2](96,72)(96,8)
    \Line[arrow,arrowpos=0.5,arrowlength=5,arrowwidth=2,arrowinset=0.2](96,8)(32,8)
    \Photon(96,8)(160,8){3}{6}
    \Text(16,4)[lb]{\Large{\Black{$\bar{e}_{i}$}}}
    \Text(94,74)[lb]{\Large{\Black{$e_{j}$}}}
    \Text(166,4)[lb]{\Large{\Black{$Z_{\mu}$}}}
  \end{picture}
}
&
\raisebox{35\unitlength}{
\begin{minipage}{5cm}
\begin{eqnarray}
&&
\frac{ie}{\sin2\Theta_{W}}\gamma_{\mu}\Big[-2\sin^{2}\Theta_{W}\delta_{ij}
+U_{L,{ja}}^{e,*}U_{L,{ia}}^{e}P_L+U_{R,{i5}}^{e,*}U_{R,{j5}}^{e}P_R\Big]\nonumber
\end{eqnarray}
\end{minipage}
}
\end{tabular}

%% fermions W
~\\
%%%%%%%%%%%%%%%%%%%%%%%%%%%%%%%%%%%%
\begin{tabular}{ll}
\fcolorbox{white}{white}{
  \begin{picture}(157,104) (19,5)
    \SetWidth{1.0}
    \SetColor{Black}
    \Line[arrow,arrowpos=0.5,arrowlength=5,arrowwidth=2,arrowinset=0.2](96,72)(96,8)
    \Line[arrow,arrowpos=0.5,arrowlength=5,arrowwidth=2,arrowinset=0.2](96,8)(32,8)
    \Photon(96,8)(160,8){3}{6}
    \Text(16,4)[lb]{\Large{\Black{$\bar{u}_{i}$}}}
    \Text(94,74)[lb]{\Large{\Black{$d_{j}$}}}
    \Text(166,4)[lb]{\Large{\Black{$W^{-}_{\mu}$}}}
  \end{picture}
}
&
\raisebox{35\unitlength}{
\begin{minipage}{5cm}
\begin{eqnarray}
&&
-i\frac{g_{2}}{\sqrt{2}}\gamma_{\mu}
\Bigl[U_{L,{ja}}^{d,*}U_{L,{ia}}^{u}P_{L}
-U_{R,{i5}}^{u,*}U_{R,{j5}}^{d}P_{R}\Bigl]\nonumber
\end{eqnarray}
\end{minipage}
}
\end{tabular}

~\\
%%%%%%%%%%%%%%%%%%%%%%%%%%%%%%%%%%%%
\begin{tabular}{ll}
\fcolorbox{white}{white}{
  \begin{picture}(157,104) (19,5)
    \SetWidth{1.0}
    \SetColor{Black}
    \Line[arrow,arrowpos=0.5,arrowlength=5,arrowwidth=2,arrowinset=0.2](96,72)(96,8)
    \Line[arrow,arrowpos=0.5,arrowlength=5,arrowwidth=2,arrowinset=0.2](96,8)(32,8)
    \Photon(96,8)(160,8){3}{6}
    \Text(16,4)[lb]{\Large{\Black{$\bar{d}_{i}$}}}
    \Text(94,74)[lb]{\Large{\Black{$u_{j}$}}}
    \Text(166,4)[lb]{\Large{\Black{$W^{+}_{\mu}$}}}
  \end{picture}
}
&
\raisebox{35\unitlength}{
\begin{minipage}{5cm}
\begin{eqnarray}
&&
-i\frac{g_{2}}{\sqrt{2}}\gamma_{\mu}
\Bigl[U_{L,{ja}}^{u,*}U_{L,{ia}}^{d}P_{L}
-U_{R,{i5}}^{d,*}U_{R,{j5}}^{u}P_{R}\Bigl]\nonumber
\end{eqnarray}
\end{minipage}
}
\end{tabular}

\subsection*{3. Sfermion-Higgs Boson}
%% sfermions Higgs
~\\
%%%%%%%%%%%%%%%%%%%%%%%%%%%%%%%%%%%%
\begin{tabular}{ll}
\fcolorbox{white}{white}{
  \begin{picture}(157,104) (19,5)
    \SetWidth{1.0}
    \SetColor{Black}
    \Line[dash,dashsize=5,arrow,arrowpos=0.5,arrowlength=5,arrowwidth=2,arrowinset=0.2](96,72)(96,8)
    \Line[dash,dashsize=5,arrow,arrowpos=0.5,arrowlength=5,arrowwidth=2,arrowinset=0.2](96,8)(32,8)
    \Line[dash,dashsize=5](96,8)(160,8)
    \Text(16,4)[lb]{\Large{\Black{$\tilde{d}_{k}^{\ast}$}}}
    \Text(94,74)[lb]{\Large{\Black{$\tilde{d}_{j}$}}}
    \Text(166,4)[lb]{\Large{\Black{$h_{i}$}}}
  \end{picture}
}
&
\raisebox{35\unitlength}{
\begin{minipage}{5cm}
\begin{eqnarray}
&&
-\frac{i}{12}\biggr\{
6\sqrt{2}\Bigl(A_{Yyd}Z_{j10}^{D,*}Z_{{k9}}^{D}
+A_{Yyd}^{*}Z_{j9}^{D,*}Z_{{k10}}^{D}\Bigr)Z_{{i2}}^{H}\nonumber\\
&&+6\sqrt{2}\Bigl(A_{d,{ab}}Z_{{k4+a}}^{D}Z_{jb}^{D,*}
+A_{d,{ab}}^{*}Z_{j4+a}^{D,*}Z_{{kb}}^{D}\Bigr)Z_{{i1}}^{H}\nonumber\\
&&
-6\sqrt{2}\Bigl(\mu^{*}Z_{{k4+a}}^{D}Y_{d,{ab}}Z_{jb}^{D,*}
+\mu Z_{j4+a}^{D,*}Y_{d,{ab}}^{*}Z_{{kb}}^{D}\Bigr)Z_{{i2}}^{H}\nonumber\\
&&
-6\sqrt{2}Y_{yd}\Bigl(\mu^{*}Z_{j10}^{D,*}Z_{{k9}}^{D}
+\mu Z_{j9}^{D,*}Z_{{k10}}^{D}\Bigr)Z_{{i1}}^{H}\nonumber\\
&&
+6\sqrt{2}\Bigl(M_{yd,{a}}Y_{d,{ab}}^{*}Z_{{kb}}^{D}Z_{j10}^{D,*}
-M_{yq,{a}}Y_{d,{ba}}Z_{{k4+b}}^{D}Z_{j9}^{D,*}\Bigr)Z_{{i1}}^{H}\nonumber\\
&&
+6\sqrt{2}\Bigl(M_{yd,{a}}Y_{d,{ab}}Z_{jb}^{D,*}Z_{{k10}}^{D}
-M_{yq,{a}}Y_{d,{ba}}^{*}Z_{j4+b}^{D,*}Z_{{k9}}^{D}\Bigr)Z_{{i1}}^{H}\nonumber\\
&&
+6\sqrt{2}Y_{yd}\Bigl(M_{yd,{a}}Z_{j4+a}^{D,*}Z_{{k9}}^{D}
-M_{yq,{a}}Z_{ja}^{D,*}Z_{{k10}}^{D}\Bigr)Z_{{i2}}^{H}\nonumber\\
&&
+6\sqrt{2}Y_{yd}\Bigl(M_{yd,{a}}Z_{j9}^{D,*}Z_{{k4+a}}^{D}-M_{yq,{a}}Z_{j10}^{D,*}Z_{{ka}}^{D}\Bigr)Z_{{i2}}^{H}\nonumber\\
&&
+12v_{u}Y_{yd}^{2}\Bigl(Z_{j9}^{D,*}Z_{{k9}}^{D}Z_{{i2}}^{H}
+Z_{j10}^{D,*}Z_{{k10}}^{D}\Bigr)Z_{{i2}}^{H}\nonumber\\
&&
+12v_{d}\Bigl(Z_{j4+c}^{D,*}Y_{d,{ca}}^{*}Y_{d,{ba}}Z_{{k4+b}}^{D}+Z_{jb}^{D,*}Y_{d,{ac}}^{*}Y_{d,{ab}}Z_{{kc}}^{D}\Bigr)Z_{{i1}}^{H}\nonumber\\
&&
+\left[\Big(3g_{2}^{2}+g_{1}^{2}\Big)Z_{j9}^{D,*}Z_{{k9}}^{D}+2g_{1}^{2}Z_{j10}^{D,*}Z_{{k10}}^{D}\right]
\left(v_{d}Z_{{i1}}^{H}-v_{u}Z_{{i2}}^{H}\right)\nonumber\\
&&
-\left[\Bigl(3g_{2}^{2}+g_{1}^{2}\Big)Z_{ja}^{D,*}Z_{{ka}}^{D}
+2g_{1}^{2}Z_{j4+a}^{D,*}Z_{{k4+a}}^{D}\right]\Big(v_{d}Z_{{i1}}^{H}-v_{u}Z_{{i2}}^{H}\Big)
\biggl\}\nonumber
\end{eqnarray}
\end{minipage}
}
\end{tabular}

~\\
%%%%%%%%%%%%%%%%%%%%%%%%%%%%%%%%%%%%
\begin{tabular}{ll}
\fcolorbox{white}{white}{
  \begin{picture}(157,104) (19,5)
    \SetWidth{1.0}
    \SetColor{Black}
    \Line[dash,dashsize=5,arrow,arrowpos=0.5,arrowlength=5,arrowwidth=2,arrowinset=0.2](96,72)(96,8)
    \Line[dash,dashsize=5,arrow,arrowpos=0.5,arrowlength=5,arrowwidth=2,arrowinset=0.2](96,8)(32,8)
    \Line[dash,dashsize=5](96,8)(160,8)
    \Text(16,4)[lb]{\Large{\Black{$\tilde{u}_{k}^{\ast}$}}}
    \Text(94,74)[lb]{\Large{\Black{$\tilde{u}_{j}$}}}
    \Text(166,4)[lb]{\Large{\Black{$h_{i}$}}}
  \end{picture}
}
&
\raisebox{35\unitlength}{
\begin{minipage}{5cm}
\begin{eqnarray}
&&
-\frac{i}{12}\biggr\{
6\sqrt{2}\Bigl(A_{Yyu}Z_{j10}^{U,*}Z_{{k9}}^{U}
+6\sqrt{2}A_{Yyu}^{*}Z_{j9}^{U,*}Z_{{k10}}^{U}\Bigr)Z_{{i1}}^{H}\nonumber\\
&&
+6\sqrt{2}\Bigl(A_{u,{ab}}Z_{{k4+a}}^{U}Z_{jb}^{U,*}
+A_{u,{ab}}^{*}Z_{j4+a}^{U,*}Z_{{kb}}^{U}\Bigr)Z_{{i2}}^{H}\nonumber\\
&&
-6\sqrt{2}\Bigl(\mu^{*}Y_{u,{ab}}Z_{jb}^{U,*}Z_{{k4+a}}^{U}
+\mu Y_{u,{ab}}^{*}Z_{j4+a}^{U,*}Z_{{kb}}^{U}\Bigr)Z_{{i1}}^{H}\nonumber\\
&&
-6\sqrt{2}Y_{yu}\Bigl(\mu Z_{j9}^{U,*}Z_{{k10}}^{U}Z_{{i2}}^{H}
+\mu^{*}Z_{j10}^{U,*}Z_{{k9}}^{U}\Bigr)Z_{{i2}}^{H}\nonumber\\
&&
+6\sqrt{2}\Bigl(M_{yq,{a}}Y_{u,{ba}}^{*}Z_{j4+b}^{U,*}Z_{{k9}}^{U}
+M_{yu,{a}}Y_{u,{ab}}^{*}Z_{j10}^{U,*}Z_{{kb}}^{U}\Bigr)Z_{{i2}}^{H}\nonumber\\
&&
+6\sqrt{2}\Bigl(M_{yq,{a}}Y_{u,{ba}}Z_{{k4+b}}^{U}Z_{j9}^{U,*}
+M_{yu,{a}}Y_{u,{ab}}Z_{jb}^{U,*}Z_{{k10}}^{U}\Bigr)Z_{{i2}}^{H}\nonumber\\
&&
+6\sqrt{2}Y_{yu}\Bigl(M_{yu,{a}}Z_{j9}^{U,*}Z_{{k4+a}}^{U}
+M_{yq,{a}}Z_{j10}^{U,*}Z_{{ka}}^{U}\Bigr)Z_{{i1}}^{H}\nonumber\\
&&
+6\sqrt{2}Y_{yu}\Bigl(M_{yu,{a}}Z_{j4+a}^{U,*}Z_{{k9}}^{U}
+M_{yq,{a}}Z_{ja}^{U,*}Z_{{k10}}^{U}\Bigr)Z_{{i1}}^{H}\nonumber\\
&&
+12v_{d}Y_{yu}^{2}\Bigl(Z_{j9}^{U,*}Z_{{k9}}^{U}
+Z_{j10}^{U,*}Z_{{k10}}^{U}\Bigr)Z_{{i1}}^{H}\nonumber\\
&&
+12v_{u}\Bigl(Z_{j4+c}^{U,*}Y_{u,{ca}}^{*}Y_{u,{ba}}Z_{{k4+b}}^{U}
+Z_{jb}^{U,*}Y_{u,{ac}}^{*}Y_{u,{ab}}Z_{{kc}}^{U}\Bigr)Z_{{i2}}^{H}\nonumber\\
&&
-\left[\left(3g_{2}^{2}-g_{1}^{2}\right)Z_{j9}^{U,*}Z_{{k9}}^{U}
+4g_{1}^{2}Z_{j10}^{U,*}Z_{{k10}}^{U}\right]\Bigl(v_{d}Z_{{i1}}^{H}-v_{u}Z_{{i2}}^{H}\Bigr)\nonumber\\
&&
+\left[\Big(3g_{2}^{2}-g_{1}^{2}\Big)Z_{ja}^{U,*}Z_{{ka}}^{U}
+4g_{1}^{2}Z_{j4+a}^{U,*}Z_{{k4+a}}^{U}\right]\Bigl(v_{d}Z_{{i1}}^{H}-v_{u}Z_{{i2}}^{H}\Bigr)\biggl\}\nonumber
\end{eqnarray}
\end{minipage}
}
\end{tabular}

~\\
%%%%%%%%%%%%%%%%%%%%%%%%%%%%%%%%%%%%
\begin{tabular}{ll}
\fcolorbox{white}{white}{
  \begin{picture}(157,104) (19,5)
    \SetWidth{1.0}
    \SetColor{Black}
    \Line[dash,dashsize=5,arrow,arrowpos=0.5,arrowlength=5,arrowwidth=2,arrowinset=0.2](96,72)(96,8)
    \Line[dash,dashsize=5,arrow,arrowpos=0.5,arrowlength=5,arrowwidth=2,arrowinset=0.2](96,8)(32,8)
    \Line[dash,dashsize=5](96,8)(160,8)
    \Text(16,4)[lb]{\Large{\Black{$\tilde{e}_{k}^{\ast}$}}}
    \Text(94,74)[lb]{\Large{\Black{$\tilde{e}_{j}$}}}
    \Text(166,4)[lb]{\Large{\Black{$h_{i}$}}}
  \end{picture}
}
&
\raisebox{35\unitlength}{
\begin{minipage}{5cm}
\begin{eqnarray}
&&
\frac{i}{4}\biggl\{
-2\sqrt{2}\Bigl(A_{Yye}Z_{j10}^{E,*}Z_{{k9}}^{E}
+A_{Yye}^{*}Z_{j9}^{E,*}Z_{{k10}}^{E}\Bigr)Z_{{i2}}^{H}\nonumber\\
&&
-2\sqrt{2}\Bigl(A_{e,{ab}}Z_{{k4+a}}^{E}Z_{jb}^{E,*}
+A_{e,{ab}}^{*}Z_{j4+a}^{E,*}Z_{{kb}}^{E}\Bigr)Z_{{i1}}^{H}\nonumber\\
&&
+2\sqrt{2}\Bigl(Z_{jb}^{E,*}\mu^{*}Y_{e,{ab}}Z_{{k4+a}}^{E}
+Z_{j4+a}^{E,*}\mu Y_{e,{ab}}^{*}Z_{{kb}}^{E}\Bigr)Z_{{i2}}^{H}\nonumber\\
&&
+2\sqrt{2}Y_{ye}\Bigl(\mu^{*}Z_{j10}^{E,*}Z_{{k9}}^{E}
+\mu Z_{j9}^{E,*}Z_{{k10}}^{E}\Bigr)Z_{{i1}}^{H}\nonumber\\
&&
+2\sqrt{2}\Bigl(M_{yl,{a}}Y_{e,{ba}}Z_{j9}^{E,*}Z_{{k4+b}}^{E}
-M_{ye,{a}}Y_{e,{ab}}^{*}Z_{j10}^{E,*}Z_{{kb}}^{E}\Bigr)Z_{{i1}}^{H}\nonumber\\
&&
+2\sqrt{2}\Bigl(M_{yl,{a}}Y_{e,{ba}}^{*}Z_{j4+b}^{E,*}Z_{{k9}}^{E}
-M_{ye,{a}}Y_{e,{ab}}Z_{jb}^{E,*}Z_{{k10}}^{E}\Bigr)Z_{{i1}}^{H}\nonumber\\
&&
+2\sqrt{2}Y_{ye}\Bigl(M_{yl,{a}}Z_{j10}^{E,*}Z_{{ka}}^{E}
-M_{ye,{a}}Z_{j9}^{E,*}Z_{{k4+a}}^{E}\Bigr)Z_{{i2}}^{H}\nonumber\\
&&
+2\sqrt{2}Y_{ye}\Bigl(Z_{ja}^{E,*}M_{yl,{a}}Z_{{k10}}^{E}
-Z_{j4+a}^{E,*}M_{ye,{a}}Z_{{k9}}^{E}\Bigr)Z_{{i2}}^{H}\nonumber\\
&&
-4v_{u}Y_{ye}^{2}\Bigl(Z_{j9}^{E,*}Z_{{k9}}^{E}
-Z_{j10}^{E,*}Z_{{k10}}^{E}\Bigr)Z_{{i2}}^{H}\nonumber\\
&&
-4v_{d}
\Bigl(Z_{j4+c}^{E,*}Y_{e,{ca}}^{*}Y_{e,{ba}}Z_{{k4+b}}^{E}
+Z_{jb}^{E,*}Y_{e,{ac}}^{*}Y_{e,{ab}}Z_{{kc}}^{E}\Bigr)Z_{{i1}}^{H}\nonumber\\
&&
+\left[\Bigl(g_{1}^{2}-g_{2}^{2}\Bigr)Z_{j9}^{E,*}Z_{{k9}}^{E}
-2g_{1}^{2}Z_{j10}^{E,*}Z_{{k10}}^{E}\right]
\left(v_{d}Z_{{i1}}^{H}-v_{u}Z_{{i2}}^{H}\right)\nonumber\\
&&
+\left[\Bigl(g_{2}^{2}-g_{1}^{2}\Bigr)\Big)Z_{ja}^{E,*}Z_{{ka}}^{E}
+2g_{1}^{2}Z_{j4+a}^{E,*}Z_{{k4+a}}^{E}\right]
\Bigl(v_{d}Z_{{i1}}^{H}-v_{u}Z_{{i2}}^{H}\Bigr)\Big)\biggr\}\nonumber
\end{eqnarray}
\end{minipage}
}
\end{tabular}

\subsection*{4. Sfermion-Gauge Boson}
%% sfermions gamma
~\\
%%%%%%%%%%%%%%%%%%%%%%%%%%%%%%%%%%%%
\begin{tabular}{ll}
\fcolorbox{white}{white}{
  \begin{picture}(157,104) (19,5)
    \SetWidth{1.0}
    \SetColor{Black}
    \Line[dash,dashsize=5,arrow,arrowpos=0.5,arrowlength=5,arrowwidth=2,arrowinset=0.2](96,72)(96,8)
    \Line[dash,dashsize=5,arrow,arrowpos=0.5,arrowlength=5,arrowwidth=2,arrowinset=0.2](96,8)(32,8)
    \Photon(96,8)(160,8){3}{6}
    \Text(16,4)[lb]{\Large{\Black{$\tilde{d}_{i}^{\ast}$}}}
    \Text(94,74)[lb]{\Large{\Black{$\tilde{d}_{j}$}}}
    \Text(166,4)[lb]{\Large{\Black{$\gamma_{\mu}$}}}
  \end{picture}
}
&
\raisebox{35\unitlength}{
\begin{minipage}{5cm}
\begin{eqnarray}
&&
-i\frac{1}{3}e\Big(Z_{ia}^{D,*}Z_{{ja}}^{D}
+Z_{i4+a}^{D,*}Z_{{j4+a}}^{D}-Z_{i9}^{D,*}Z_{{j9}}^{D}
-Z_{i10}^{D,*}Z_{{j10}}^{D}\Big)
\Big(p_{\mu}^{\tilde{d}_{{i}}^{*}}-p_{\mu}^{\tilde{d}_{{j}}}\Big)\nonumber
\end{eqnarray}
\end{minipage}
}
\end{tabular}

~\\
%%%%%%%%%%%%%%%%%%%%%%%%%%%%%%%%%%%%
\begin{tabular}{ll}
\fcolorbox{white}{white}{
  \begin{picture}(157,104) (19,5)
    \SetWidth{1.0}
    \SetColor{Black}
    \Line[dash,dashsize=5,arrow,arrowpos=0.5,arrowlength=5,arrowwidth=2,arrowinset=0.2](96,72)(96,8)
    \Line[dash,dashsize=5,arrow,arrowpos=0.5,arrowlength=5,arrowwidth=2,arrowinset=0.2](96,8)(32,8)
    \Photon(96,8)(160,8){3}{6}
    \Text(16,4)[lb]{\Large{\Black{$\tilde{u}_{i}^{\ast}$}}}
    \Text(94,74)[lb]{\Large{\Black{$\tilde{u}_{j}$}}}
    \Text(166,4)[lb]{\Large{\Black{$\gamma_{\mu}$}}}
  \end{picture}
}
&
\raisebox{35\unitlength}{
\begin{minipage}{5cm}
\begin{eqnarray}
&&
i\frac{2}{3}e\Big(Z_{ia}^{U,*}Z_{{ja}}^{U}
+Z_{i4+a}^{U,*}Z_{{j4+a}}^{U}
-Z_{i9}^{U,*}Z_{{j9}}^{U}-Z_{i10}^{U,*}Z_{{j10}}^{U}\Big)
\Big(p_{\mu}^{\tilde{u}_{{i}}^{*}}-p_{\mu}^{\tilde{u}_{{j}}}\Big)\nonumber
\end{eqnarray}
\end{minipage}
}
\end{tabular}

~\\
%%%%%%%%%%%%%%%%%%%%%%%%%%%%%%%%%%%%
\begin{tabular}{ll}
\fcolorbox{white}{white}{
  \begin{picture}(157,104) (19,5)
    \SetWidth{1.0}
    \SetColor{Black}
    \Line[dash,dashsize=5,arrow,arrowpos=0.5,arrowlength=5,arrowwidth=2,arrowinset=0.2](96,72)(96,8)
    \Line[dash,dashsize=5,arrow,arrowpos=0.5,arrowlength=5,arrowwidth=2,arrowinset=0.2](96,8)(32,8)
    \Photon(96,8)(160,8){3}{6}
    \Text(16,4)[lb]{\Large{\Black{$\tilde{e}_{i}^{\ast}$}}}
    \Text(94,74)[lb]{\Large{\Black{$\tilde{e}_{j}$}}}
    \Text(166,4)[lb]{\Large{\Black{$\gamma_{\mu}$}}}
  \end{picture}
}
&
\raisebox{35\unitlength}{
\begin{minipage}{5cm}
\begin{eqnarray}
&&
-ie\Big(Z_{ia}^{E,*}Z_{{ja}}^{E}
+Z_{i4+a}^{E,*}Z_{{j4+a}}^{E}
-Z_{i9}^{E,*}Z_{{j9}}^{E}-Z_{i10}^{E,*}Z_{{j10}}^{E}\Big)
\Big(p_{\mu}^{\tilde{e}_{{i}}^{*}}-p_{\mu}^{\tilde{e}_{{j}}}\Big)\nonumber
\end{eqnarray}
\end{minipage}
}
\end{tabular}

%% sfermions Z
~\\
%%%%%%%%%%%%%%%%%%%%%%%%%%%%%%%%%%%%
\begin{tabular}{ll}
\fcolorbox{white}{white}{
  \begin{picture}(157,104) (19,5)
    \SetWidth{1.0}
    \SetColor{Black}
    \Line[dash,dashsize=5,arrow,arrowpos=0.5,arrowlength=5,arrowwidth=2,arrowinset=0.2](96,72)(96,8)
    \Line[dash,dashsize=5,arrow,arrowpos=0.5,arrowlength=5,arrowwidth=2,arrowinset=0.2](96,8)(32,8)
    \Photon(96,8)(160,8){3}{6}
    \Text(16,4)[lb]{\Large{\Black{$\tilde{d}_{i}^{\ast}$}}}
    \Text(94,74)[lb]{\Large{\Black{$\tilde{d}_{j}$}}}
    \Text(166,4)[lb]{\Large{\Black{$Z_{\mu}$}}}
  \end{picture}
}
&
\raisebox{35\unitlength}{
\begin{minipage}{5cm}
\begin{eqnarray}
&&
\frac{ie}{3\sin2\Theta_{W}}\Big(\Big(3-2\sin^{2}\Theta_{W}\Big)
Z_{ia}^{D,*}Z_{{ja}}^{D}-2\sin^{2}\Theta_{W}
Z_{i4+a}^{D,*}Z_{{j4+a}}^{D}\nonumber\\
&&
-\Big(3-2\sin^{2}\Theta_{W}\Big)Z_{i9}^{D,*}Z_{{j9}}^{D}
+2\sin^{2}\Theta_{W}Z_{i10}^{D,*}Z_{{j10}}^{D}\Big)
\Big(p_{\mu}^{\tilde{d}_{{i}}^{*}}
-p_{\mu}^{\tilde{d}_{{j}}}\Big)\nonumber
\end{eqnarray}
\end{minipage}
}
\end{tabular}

~\\
%%%%%%%%%%%%%%%%%%%%%%%%%%%%%%%%%%%%
\begin{tabular}{ll}
\fcolorbox{white}{white}{
  \begin{picture}(157,104) (19,5)
    \SetWidth{1.0}
    \SetColor{Black}
    \Line[dash,dashsize=5,arrow,arrowpos=0.5,arrowlength=5,arrowwidth=2,arrowinset=0.2](96,72)(96,8)
    \Line[dash,dashsize=5,arrow,arrowpos=0.5,arrowlength=5,arrowwidth=2,arrowinset=0.2](96,8)(32,8)
    \Photon(96,8)(160,8){3}{6}
    \Text(16,4)[lb]{\Large{\Black{$\tilde{u}_{i}^{\ast}$}}}
    \Text(94,74)[lb]{\Large{\Black{$\tilde{u}_{j}$}}}
    \Text(166,4)[lb]{\Large{\Black{$Z_{\mu}$}}}
  \end{picture}
}
&
\raisebox{35\unitlength}{
\begin{minipage}{5cm}
\begin{eqnarray}
&&
-\frac{ie}{3\sin^{2}\Theta_{W}}\Big(\Big(3-4\sin^{2}\Theta_{W}\Big)
Z_{ia}^{U,*}Z_{{ja}}^{U}-4\sin^{2}\Theta_{W}
Z_{i4+a}^{U,*}Z_{{j4+a}}^{U}\nonumber\\
&&
-\Big(3-4\sin^{2}\Theta_{W}\Big)Z_{i9}^{U,*}Z_{{j9}}^{U}
+4\sin^{2}\Theta_{W}Z_{i10}^{U,*}Z_{{j10}}^{U}\Big)
\Big(p_{\mu}^{\tilde{u}_{{i}}^{*}}
-p_{\mu}^{\tilde{u}_{{j}}}\Big)\nonumber
\end{eqnarray}
\end{minipage}
}
\end{tabular}

~\\
%%%%%%%%%%%%%%%%%%%%%%%%%%%%%%%%%%%%
\begin{tabular}{ll}
\fcolorbox{white}{white}{
  \begin{picture}(157,104) (19,5)
    \SetWidth{1.0}
    \SetColor{Black}
    \Line[dash,dashsize=5,arrow,arrowpos=0.5,arrowlength=5,arrowwidth=2,arrowinset=0.2](96,72)(96,8)
    \Line[dash,dashsize=5,arrow,arrowpos=0.5,arrowlength=5,arrowwidth=2,arrowinset=0.2](96,8)(32,8)
    \Photon(96,8)(160,8){3}{6}
    \Text(16,4)[lb]{\Large{\Black{$\tilde{e}_{i}^{\ast}$}}}
    \Text(94,74)[lb]{\Large{\Black{$\tilde{e}_{j}$}}}
    \Text(166,4)[lb]{\Large{\Black{$Z_{\mu}$}}}
  \end{picture}
}
&
\raisebox{35\unitlength}{
\begin{minipage}{5cm}
\begin{eqnarray}
&&
\frac{ie}{\sin2\Theta_{W}}\Big(\Big(1-2\sin^{2}\Theta_{W}\Big)
Z_{ia}^{E,*}Z_{{ja}}^{E}-2\sin^{2}\Theta_{W}
Z_{i4+a}^{E,*}Z_{{j4+a}}^{E}\nonumber\\
&&
-\Bigl(1-2\sin^{2}\Theta_{W}\Bigr)Z_{i9}^{E,*}Z_{{j9}}^{E}
+2\sin^{2}\Theta_{W}Z_{i10}^{E,*}Z_{{j10}}^{E}\Big)
\Big(p_{\mu}^{\tilde{e}_{{i}}^{*}}
-p_{\mu}^{\tilde{e}_{{j}}}\Big)\nonumber
\end{eqnarray}
\end{minipage}
}
\end{tabular}

%% sfermions W
~\\
%%%%%%%%%%%%%%%%%%%%%%%%%%%%%%%%%%%%
\begin{tabular}{ll}
\fcolorbox{white}{white}{
  \begin{picture}(157,104) (19,5)
    \SetWidth{1.0}
    \SetColor{Black}
    \Line[dash,dashsize=5,arrow,arrowpos=0.5,arrowlength=5,arrowwidth=2,arrowinset=0.2](96,72)(96,8)
    \Line[dash,dashsize=5,arrow,arrowpos=0.5,arrowlength=5,arrowwidth=2,arrowinset=0.2](96,8)(32,8)
    \Photon(96,8)(160,8){3}{6}
    \Text(16,4)[lb]{\Large{\Black{$\tilde{u}_{i}^{\ast}$}}}
    \Text(94,74)[lb]{\Large{\Black{$\tilde{d}_{j}$}}}
    \Text(166,4)[lb]{\Large{\Black{$W^{-}_{\mu}$}}}
  \end{picture}
}
&
\raisebox{35\unitlength}{
\begin{minipage}{5cm}
\begin{eqnarray}
&&
-i\frac{1}{\sqrt{2}}g_{2}\left(Z_{ia}^{U,*}Z_{{ja}}^{D}
+Z_{i9}^{U,*}Z_{{j9}}^{D}\right)\Big(p_{\mu}^{\tilde{u}_{{i}}^{*}}
-p_{\mu}^{\tilde{d}_{{j}}}\Big)\nonumber
\end{eqnarray}
\end{minipage}
}
\end{tabular}

~\\
%%%%%%%%%%%%%%%%%%%%%%%%%%%%%%%%%%%%
\begin{tabular}{ll}
\fcolorbox{white}{white}{
  \begin{picture}(157,104) (19,5)
    \SetWidth{1.0}
    \SetColor{Black}
    \Line[dash,dashsize=5,arrow,arrowpos=0.5,arrowlength=5,arrowwidth=2,arrowinset=0.2](96,72)(96,8)
    \Line[dash,dashsize=5,arrow,arrowpos=0.5,arrowlength=5,arrowwidth=2,arrowinset=0.2](96,8)(32,8)
    \Photon(96,8)(160,8){3}{6}
    \Text(16,4)[lb]{\Large{\Black{$\tilde{d}_{i}^{\ast}$}}}
    \Text(94,74)[lb]{\Large{\Black{$\tilde{u}_{j}$}}}
    \Text(166,4)[lb]{\Large{\Black{$W^{+}_{\mu}$}}}
  \end{picture}
}
&
\raisebox{35\unitlength}{
\begin{minipage}{5cm}
\begin{eqnarray}
&&
-i\frac{1}{\sqrt{2}}g_{2}\left(Z_{ia}^{D,*}Z_{{ja}}^{U}
+Z_{i9}^{D,*}Z_{{j9}}^{U}\right)\Big(p_{\mu}^{\tilde{d}_{{i}}^{*}}
-p_{\mu}^{\tilde{u}_{{j}}}\Big)\nonumber
\end{eqnarray}
\end{minipage}
}
\end{tabular}

\subsection*{5. Chargino/Neutralino-Fermion and Sfermion}
%% sfermions \chi
~\\
%%%%%%%%%%%%%%%%%%%%%%%%%%%%%%%%%%%%
\begin{tabular}{ll}
\fcolorbox{white}{white}{
  \begin{picture}(157,104) (19,5)
    \SetWidth{1.0}
    \SetColor{Black}
    \Line[arrow,arrowpos=0.5,arrowlength=5,arrowwidth=2,arrowinset=0.2](96,72)(96,8)
    \Line[arrow,arrowpos=0.5,arrowlength=5,arrowwidth=2,arrowinset=0.2](96,8)(32,8)
    \Line[dash,dashsize=5,arrow,arrowpos=0.5,arrowlength=5,arrowwidth=2,arrowinset=0.2](96,8)(160,8)
    \Text(16,4)[lb]{\Large{\Black{$\tilde{\chi}_{i}^{-}$}}}
    \Text(94,74)[lb]{\Large{\Black{$u_{j}$}}}
    \Text(166,4)[lb]{\Large{\Black{$\tilde{d}_{k}^{\ast}$}}}
  \end{picture}
}
&
\raisebox{35\unitlength}{
\begin{minipage}{5cm}
\begin{eqnarray}
&&
-i\Big(g_{2}U_{i1}^{*}U_{L,{ja}}^{u,*}Z_{{ka}}^{D}
-Y_{yu}U_{i2}^{*}U_{L,{j5}}^{u,*}Z_{{k9}}^{D}
-\sum_{a=1}^{4}Y_{d,{ab}}U_{i2}^{*}U_{L,{jb}}^{u,*}Z_{{k4+a}}^{D}\Big)P_{L}\nonumber\\
&&
-i\Big(g_{2}V_{{i1}}U_{R,{j5}}^{u}Z_{{k9}}^{D}
-Y_{yd}V_{{i2}}U_{R,{j5}}^{u}Z_{{k10}}^{D}
-sum_{b=1}^{4}Y_{u,{ab}}^{*}V_{{i2}}U_{R,{ja}}^{u}Z_{{kb}}^{D}\Big)P_{R}\nonumber
\end{eqnarray}
\end{minipage}
}
\end{tabular}

~\\
%%%%%%%%%%%%%%%%%%%%%%%%%%%%%%%%%%%%
\begin{tabular}{ll}
\fcolorbox{white}{white}{
  \begin{picture}(157,104) (19,5)
    \SetWidth{1.0}
    \SetColor{Black}
    \Line[arrow,arrowpos=0.5,arrowlength=5,arrowwidth=2,arrowinset=0.2](96,72)(96,8)
    \Line[arrow,arrowpos=0.5,arrowlength=5,arrowwidth=2,arrowinset=0.2](96,8)(32,8)
    \Line[dash,dashsize=5,arrow,arrowpos=0.5,arrowlength=5,arrowwidth=2,arrowinset=0.2](96,8)(160,8)
    \Text(16,4)[lb]{\Large{\Black{$\tilde{\chi}_{i}^{-}$}}}
    \Text(94,74)[lb]{\Large{\Black{$\bar{d}_{j}$}}}
    \Text(166,4)[lb]{\Large{\Black{$\tilde{u}_{k}$}}}
  \end{picture}
}
&
\raisebox{35\unitlength}{
\begin{minipage}{5cm}
\begin{eqnarray}
&&
-i\Big(g_{2}U_{R,{j5}}^{d,*}Z_{k9}^{U,*}U_{i1}^{*}
-Y_{yu}U_{R,{j5}}^{d,*}Z_{k10}^{U,*}U_{i2}^{*}
-\sum_{b=1}^{4}Y_{d,{ab}}U_{R,{ja}}^{d,*}Z_{kb}^{U,*}U_{i2}^{*}\Big)P_{L}\nonumber\\
&&
-i\Big(g_{2}U_{L,{ja}}^{d}Z_{ka}^{U,*}V_{{i1}}
-Y_{yd}U_{L,{j5}}^{d}Z_{k9}^{U,*}V_{{i2}}
-\sum_{a=1}^{4}Y_{u,{ab}}^{*}U_{L,{jb}}^{d}Z_{k4+a}^{U,*}V_{{i2}}\Big)P_{R}\nonumber
\end{eqnarray}
\end{minipage}
}
\end{tabular}

~\\
%%%%%%%%%%%%%%%%%%%%%%%%%%%%%%%%%%%%
\begin{tabular}{ll}
\fcolorbox{white}{white}{
  \begin{picture}(157,104) (19,5)
    \SetWidth{1.0}
    \SetColor{Black}
    \Line[arrow,arrowpos=0.5,arrowlength=5,arrowwidth=2,arrowinset=0.2](96,72)(96,8)
    \Line[arrow,arrowpos=0.5,arrowlength=5,arrowwidth=2,arrowinset=0.2](96,8)(32,8)
    \Line[dash,dashsize=5,arrow,arrowpos=0.5,arrowlength=5,arrowwidth=2,arrowinset=0.2](96,8)(160,8)
    \Text(16,4)[lb]{\Large{\Black{$d_{j}$}}}
    \Text(94,74)[lb]{\Large{\Black{$\tilde{\chi}_{i}^{+}$}}}
    \Text(166,4)[lb]{\Large{\Black{$\tilde{u}_{k}^{\ast}$}}}
  \end{picture}
}
&
\raisebox{35\unitlength}{
\begin{minipage}{5cm}
\begin{eqnarray}
&&
-i\Big(g_{2}Z_{{ka}}^{U}U_{L,{ja}}^{d,*}V_{i1}^{*}
-Y_{yd}Z_{{k9}}^{U}U_{L,{j5}}^{d,*}V_{i2}^{*}-
Z_{{k4+a}}^{U}U_{L,{jb}}^{d,*}Y_{u,{ab}}V_{i2}^{*}\Big)P_{L}\nonumber\\
&&
-i\Big(g_{2}Z_{{k9}}^{U}U_{R,{j5}}^{d}U_{{i1}}
-Y_{yu}Z_{{k10}}^{U}U_{R,{j5}}^{d}U_{{i2}}
-Z_{{kb}}^{U}U_{R,{ja}}^{d}Y_{d,{ab}}^{*}U_{{i2}}\Big)P_{R}\nonumber
\end{eqnarray}
\end{minipage}
}
\end{tabular}

~\\
%%%%%%%%%%%%%%%%%%%%%%%%%%%%%%%%%%%%
\begin{tabular}{ll}
\fcolorbox{white}{white}{
  \begin{picture}(157,104) (19,5)
    \SetWidth{1.0}
    \SetColor{Black}
    \Line[arrow,arrowpos=0.5,arrowlength=5,arrowwidth=2,arrowinset=0.2](96,72)(96,8)
    \Line[arrow,arrowpos=0.5,arrowlength=5,arrowwidth=2,arrowinset=0.2](96,8)(32,8)
    \Line[dash,dashsize=5,arrow,arrowpos=0.5,arrowlength=5,arrowwidth=2,arrowinset=0.2](96,8)(160,8)
    \Text(16,4)[lb]{\Large{\Black{$\bar{u}_{j}$}}}
    \Text(94,74)[lb]{\Large{\Black{$\tilde{\chi}_{i}^{+}$}}}
    \Text(166,4)[lb]{\Large{\Black{$\tilde{d}_{k}$}}}
  \end{picture}
}
&
\raisebox{35\unitlength}{
\begin{minipage}{5cm}
\begin{eqnarray}
&&
-i\Big(g_{2}Z_{k9}^{D,*}U_{R,{j5}}^{u,*}V_{i1}^{*}
-Y_{yd}Z_{k10}^{D,*}U_{R,{j5}}^{u,*}V_{i2}^{*}
-Z_{kb}^{D,*}U_{R,{ja}}^{u,*}Y_{u,{ab}}V_{i2}^{*}\Big)P_{L}\nonumber\\
&&
-i\Big(g_{2}Z_{ka}^{D,*}U_{L,{ja}}^{u}U_{{i1}}
-Y_{yu}Z_{k9}^{D,*}U_{L,{j5}}^{u}U_{{i2}}
-Z_{k4+a}^{D,*}U_{L,{jb}}^{u}Y_{d,{ab}}^{*}U_{{i2}}\Big)P_{R}\nonumber
\end{eqnarray}
\end{minipage}
}
\end{tabular}

~\\
%%%%%%%%%%%%%%%%%%%%%%%%%%%%%%%%%%%%
\begin{tabular}{ll}
\fcolorbox{white}{white}{
  \begin{picture}(157,104) (19,5)
    \SetWidth{1.0}
    \SetColor{Black}
    \Line[arrow,arrowpos=0.5,arrowlength=5,arrowwidth=2,arrowinset=0.2](96,72)(96,8)
    \Line[arrow,arrowpos=0.5,arrowlength=5,arrowwidth=2,arrowinset=0.2](96,8)(32,8)
    \Line[dash,dashsize=5,arrow,arrowpos=0.5,arrowlength=5,arrowwidth=2,arrowinset=0.2](96,8)(160,8)
    \Text(16,4)[lb]{\Large{\Black{$\tilde{\chi}^{0}$}}}
    \Text(94,74)[lb]{\Large{\Black{$d_{j}$}}}
    \Text(166,4)[lb]{\Large{\Black{$\tilde{d}_{k}^{\ast}$}}}
  \end{picture}
}
&
\raisebox{35\unitlength}{
\begin{minipage}{5cm}
\begin{eqnarray}
&&
\frac{i}{6}\Big(3\sqrt{2}g_{2}N_{i2}^{*}U_{L,{ja}}^{d,*}Z_{{ka}}^{D}
-6N_{i3}^{*}U_{L,{jb}}^{d,*}Y_{d,{ab}}Z_{{k4+a}}^{D}\nonumber\\
&&
-6Y_{yd}U_{L,{j5}}^{d,*}N_{i4}^{*}Z_{{k9}}^{D}
+2\sqrt{2}g_{1}N_{i1}^{*}U_{L,{j5}}^{d,*}Z_{{k10}}^{D}
-\sqrt{2}g_{1}N_{i1}^{*}U_{L,{ja}}^{d,*}Z_{{ka}}^{D}\Big)P_{L}\nonumber\\
&&
-\frac{i}{6}\Big(3\sqrt{2}g_{2}Z_{{k9}}^{D}U_{R,{j5}}^{d}N_{{i2}}
+6Y_{d,{ab}}^{*}U_{R,{ja}}^{d}Z_{{kb}}^{D}N_{{i3}}\nonumber\\
&&
+6Y_{yd}Z_{{k10}}^{D}U_{R,{j5}}^{d}N_{{i4}}
+2\sqrt{2}g_{1}\sum_{a=1}^{4}Z_{{k4+a}}^{D}U_{R,{ja}}^{d}N_{{i1}}
-\sqrt{2}g_{1}Z_{{k9}}^{D}U_{R,{j5}}^{d}N_{{i1}}\Big)P_{R}\nonumber
\end{eqnarray}
\end{minipage}
}
\end{tabular}

~\\
%%%%%%%%%%%%%%%%%%%%%%%%%%%%%%%%%%%%
\begin{tabular}{ll}
\fcolorbox{white}{white}{
  \begin{picture}(157,104) (19,5)
    \SetWidth{1.0}
    \SetColor{Black}
    \Line[arrow,arrowpos=0.5,arrowlength=5,arrowwidth=2,arrowinset=0.2](96,72)(96,8)
    \Line[arrow,arrowpos=0.5,arrowlength=5,arrowwidth=2,arrowinset=0.2](96,8)(32,8)
    \Line[dash,dashsize=5,arrow,arrowpos=0.5,arrowlength=5,arrowwidth=2,arrowinset=0.2](96,8)(160,8)
    \Text(16,4)[lb]{\Large{\Black{$\tilde{\chi}^{0}$}}}
    \Text(94,74)[lb]{\Large{\Black{$u_{j}$}}}
    \Text(166,4)[lb]{\Large{\Black{$\tilde{u}_{k}^{\ast}$}}}
  \end{picture}
}
&
\raisebox{35\unitlength}{
\begin{minipage}{5cm}
\begin{eqnarray}
&&
-\frac{i}{6}\Big(3\sqrt{2}g_{2}N_{i2}^{*}U_{L,{ja}}^{u,*}Z_{{ka}}^{U}
+6N_{i4}^{*}U_{L,{jb}}^{u,*}Y_{u,{ab}}Z_{{k4+a}}^{U}\nonumber\\
&&
+6Y_{yu}N_{i3}^{*}U_{L,{j5}}^{u,*}Z_{{k9}}^{U}
+4\sqrt{2}g_{1}N_{i1}^{*}U_{L,{j5}}^{u,*}Z_{{k10}}^{U}
+\sqrt{2}g_{1}N_{i1}^{*}U_{L,{ja}}^{u,*}Z_{{ka}}^{U}\Big)P_{L}\nonumber\\
&&
+\frac{i}{6}\Big(3\sqrt{2}g_{2}N_{{i2}}Z_{{k9}}^{U}U_{R,{j5}}^{u}
-6Y_{u,{ab}}^{*}U_{R,{ja}}^{u}Z_{{kb}}^{U}N_{{i4}}\nonumber\\
&&
-6Y_{yu}N_{{i3}}Z_{{k10}}^{U}U_{R,{j5}}^{u}
+4\sqrt{2}g_{1}Z_{{k4+a}}^{U}U_{R,{ja}}^{u}N_{{i1}}
+\sqrt{2}g_{1}N_{{i1}}Z_{{k9}}^{U}U_{R,{j5}}^{u}\Big)P_{R}\nonumber
\end{eqnarray}
\end{minipage}
}
\end{tabular}

~\\
%%%%%%%%%%%%%%%%%%%%%%%%%%%%%%%%%%%%
\begin{tabular}{ll}
\fcolorbox{white}{white}{
  \begin{picture}(157,104) (19,5)
    \SetWidth{1.0}
    \SetColor{Black}
    \Line[arrow,arrowpos=0.5,arrowlength=5,arrowwidth=2,arrowinset=0.2](96,72)(96,8)
    \Line[arrow,arrowpos=0.5,arrowlength=5,arrowwidth=2,arrowinset=0.2](96,8)(32,8)
    \Line[dash,dashsize=5,arrow,arrowpos=0.5,arrowlength=5,arrowwidth=2,arrowinset=0.2](96,8)(160,8)
    \Text(16,4)[lb]{\Large{\Black{$\tilde{\chi}^{0}$}}}
    \Text(94,74)[lb]{\Large{\Black{$e_{j}$}}}
    \Text(166,4)[lb]{\Large{\Black{$\tilde{e}_{k}^{\ast}$}}}
  \end{picture}
}
&
\raisebox{35\unitlength}{
\begin{minipage}{5cm}
\begin{eqnarray}
&&
\frac{i}{2}\Big(\sqrt{2}g_{2}N_{i2}^{*}U_{L,{ja}}^{e,*}Z_{{ka}}^{E}
-2N_{i3}^{*}U_{L,{jb}}^{e,*}Y_{e,{ab}}Z_{{k4+a}}^{E}\nonumber\\
&&
-2Y_{ye}N_{i4}^{*}U_{L,{j5}}^{e,*}Z_{{k9}}^{E}
+2\sqrt{2}g_{1}N_{i1}^{*}U_{L,{j5}}^{e,*}Z_{{k10}}^{E}
+\sqrt{2}g_{1}N_{i1}^{*}U_{L,{ja}}^{e,*}Z_{{ka}}^{E}\Big)P_{L}\nonumber\\
&&
-\frac{i}{2}\Big(\sqrt{2}g_{2}Z_{{k9}}^{E}U_{R,{j5}}^{e}N_{{i2}}
+2Y_{e,{ab}}^{*}U_{R,{ja}}^{e}Z_{{kb}}^{E}N_{{i3}}\nonumber\\
&&
+2Y_{ye}^{*}Z_{{k10}}^{E}U_{R,{j5}}^{e}N_{{i4}}
+\sqrt{2}g_{1}Z_{{k9}}^{E}U_{R,{j5}}^{e}N_{{i1}}
+2\sqrt{2}g_{1}Z_{{k4+a}}^{E}U_{R,{ja}}^{e}N_{{i1}}\Big)P_{R}\nonumber
\end{eqnarray}
\end{minipage}
}
\end{tabular}

~\\
%%%%%%%%%%%%%%%%%%%%%%%%%%%%%%%%%%%%
\begin{tabular}{ll}
\fcolorbox{white}{white}{
  \begin{picture}(157,104) (19,5)
    \SetWidth{1.0}
    \SetColor{Black}
    \Line[arrow,arrowpos=0.5,arrowlength=5,arrowwidth=2,arrowinset=0.2](96,72)(96,8)
    \Line[arrow,arrowpos=0.5,arrowlength=5,arrowwidth=2,arrowinset=0.2](96,8)(32,8)
    \Line[dash,dashsize=5,arrow,arrowpos=0.5,arrowlength=5,arrowwidth=2,arrowinset=0.2](96,8)(160,8)
    \Text(16,4)[lb]{\Large{\Black{$\bar{d}_{i}$}}}
    \Text(94,74)[lb]{\Large{\Black{$\tilde{\chi}_{j}^{0}$}}}
    \Text(166,4)[lb]{\Large{\Black{$\tilde{d}_{k}^{\ast}$}}}
  \end{picture}
}
&
\raisebox{35\unitlength}{
\begin{minipage}{5cm}
\begin{eqnarray}
&&
-\frac{i}{6}\Big(6Z_{kb}^{D,*}Y_{d,{ab}}U_{R,{ia}}^{d,*}N_{j3}^{*}
+6Y_{yd}Z_{k10}^{D,*}U_{R,{i5}}^{d,*}N_{j4}^{*}\nonumber\\
&&
+3\sqrt{2}g_{2}Z_{k9}^{D,*}U_{R,{i5}}^{d,*}N_{j2}^{*}
-\sqrt{2}g_{1}Z_{k9}^{D,*}U_{R,{i5}}^{d,*}N_{j1}^{*}
+2\sqrt{2}g_{1}Z_{k4+a}^{D,*}U_{R,{ia}}^{d,*}N_{j1}^{*}\Big)P_{L}\nonumber\\
&&
-\frac{i}{6}\Big(6Z_{k4+a}^{D,*}Y_{d,{ab}}^{*}U_{L,{ib}}^{d}N_{{j3}}
+6Y_{yd}Z_{k9}^{D,*}U_{L,{i5}}^{d}N_{{j4}}\nonumber\\
&&
-3\sqrt{2}g_{2}Z_{ka}^{D,*}U_{L,{ia}}^{d}N_{{j2}}
+\sqrt{2}g_{1}Z_{ka}^{D,*}U_{L,{ia}}^{d}N_{{j1}}
-2\sqrt{2}g_{1}Z_{k10}^{D,*}U_{L,{i5}}^{d}N_{{j1}}\Big)P_{R}\nonumber
\end{eqnarray}
\end{minipage}
}
\end{tabular}

~\\
%%%%%%%%%%%%%%%%%%%%%%%%%%%%%%%%%%%%
\begin{tabular}{ll}
\fcolorbox{white}{white}{
  \begin{picture}(157,104) (19,5)
    \SetWidth{1.0}
    \SetColor{Black}
    \Line[arrow,arrowpos=0.5,arrowlength=5,arrowwidth=2,arrowinset=0.2](96,72)(96,8)
    \Line[arrow,arrowpos=0.5,arrowlength=5,arrowwidth=2,arrowinset=0.2](96,8)(32,8)
    \Line[dash,dashsize=5,arrow,arrowpos=0.5,arrowlength=5,arrowwidth=2,arrowinset=0.2](96,8)(160,8)
    \Text(16,4)[lb]{\Large{\Black{$\bar{u}_{i}$}}}
    \Text(94,74)[lb]{\Large{\Black{$\tilde{\chi}_{j}^{0}$}}}
    \Text(166,4)[lb]{\Large{\Black{$\tilde{u}_{k}^{\ast}$}}}
  \end{picture}
}
&
\raisebox{35\unitlength}{
\begin{minipage}{5cm}
\begin{eqnarray}
&&
-\frac{i}{6}\Big(6U_{R,{ia}}^{u,*}Y_{u,{ab}}Z_{kb}^{U,*}N_{j4}^{*}
-4\sqrt{2}g_{1}Z_{k4+a}^{U,*}U_{R,{ia}}^{u,*}N_{j1}^{*}\nonumber\\
&&
-\sqrt{2}g_{1}Z_{k9}^{U,*}U_{R,{i5}}^{u,*}N_{j1}^{*}
-3\sqrt{2}g_{2}Z_{k9}^{U,*}U_{R,{i5}}^{u,*}N_{j2}^{*}
+6Y_{yu}Z_{k10}^{U,*}U_{R,{i5}}^{u,*}N_{j3}^{*}\Big)P_{L}\nonumber\\
&&
-\frac{i}{6}\Big(6Z_{k4+a}^{U,*}Y_{u,{ab}}^{*}U_{L,{ib}}^{u}N_{{j4}}
+4\sqrt{2}g_{1}Z_{k10}^{U,*}U_{L,{i5}}^{u}N_{{j1}}\nonumber\\
&&
+\sqrt{2}g_{1}Z_{ka}^{U,*}U_{L,{ia}}^{u}N_{{j1}}
+3\sqrt{2}g_{2}Z_{ka}^{U,*}U_{L,{ia}}^{u}N_{{j2}}
+6Y_{yu}Z_{k9}^{U,*}U_{L,{i5}}^{u}N_{{j3}}\Big)P_{R}\nonumber
\end{eqnarray}
\end{minipage}
}
\end{tabular}

~\\
%%%%%%%%%%%%%%%%%%%%%%%%%%%%%%%%%%%%
\begin{tabular}{ll}
\fcolorbox{white}{white}{
  \begin{picture}(157,104) (19,5)
    \SetWidth{1.0}
    \SetColor{Black}
    \Line[arrow,arrowpos=0.5,arrowlength=5,arrowwidth=2,arrowinset=0.2](96,72)(96,8)
    \Line[arrow,arrowpos=0.5,arrowlength=5,arrowwidth=2,arrowinset=0.2](96,8)(32,8)
    \Line[dash,dashsize=5,arrow,arrowpos=0.5,arrowlength=5,arrowwidth=2,arrowinset=0.2](96,8)(160,8)
    \Text(16,4)[lb]{\Large{\Black{$\bar{e}_{i}$}}}
    \Text(94,74)[lb]{\Large{\Black{$\tilde{\chi}_{j}^{0}$}}}
    \Text(166,4)[lb]{\Large{\Black{$\tilde{e}_{k}^{\ast}$}}}
  \end{picture}
}
&
\raisebox{35\unitlength}{
\begin{minipage}{5cm}
\begin{eqnarray}
&&
-\frac{i}{2}\Big(2\sqrt{2}g_{1}Z_{k4+a}^{E,*}U_{R,{ia}}^{e,*}N_{j1}^{*}
+2\sum_{a=1}^{4}Z_{kb}^{E,*}U_{R,{ia}}^{e,*}Y_{e,{ab}}N_{j3}^{*}\nonumber\\
&&
+\sqrt{2}g_{1}Z_{k9}^{E,*}U_{R,{i5}}^{e,*}N_{j1}^{*}
+\sqrt{2}g_{2}Z_{k9}^{E,*}U_{R,{i5}}^{e,*}N_{j2}^{*}
+2Y_{ye}Z_{k10}^{E,*}U_{R,{i5}}^{e,*}N_{j4}^{*}\Big)P_{L}\nonumber\\
&&
+\frac{i}{2}\Big(2\sqrt{2}g_{1}Z_{k10}^{E,*}U_{L,{i5}}^{e}N_{{j1}}
-2Z_{k4+a}^{E,*}U_{L,{ib}}^{e}Y_{e,{ab}}^{*}N_{{j3}}\nonumber\\
&&
+\sqrt{2}g_{1}Z_{ka}^{E,*}U_{L,{ia}}^{e}N_{{j1}}
+\sqrt{2}g_{2}Z_{ka}^{E,*}U_{L,{ia}}^{e}N_{{j2}}
-2Y_{ye}^{*}Z_{k9}^{E,*}U_{L,{i5}}^{e}N_{{j4}}\Big)P_{R}\nonumber
\end{eqnarray}
\end{minipage}
}
\end{tabular}

%%%%%%%%%%%%%%%%%%%%%%%%%%%%%%%%%%%%%%%%%%%%%%%%%%%%%%%%
%%%%%%%%%%%%%%%%%%%%%%%%%%%%%%%%%%%%%%%%%%%%%%%%%%%%%%%%%%%%%%%%%%%%%%%%%%%%%%%%%%%%
%~\\
%\clearpage

%%%%%%%%%%%%%%%%%%%%%%%%%%%%%%%%%%%%%%%%%%%%%%%% %%%%%%%%%%%%%%%%%%%%%%%%%%%%%%%%%%%%%%%%%%%%%%%%
%
\end{document}